\documentclass[12pt,preprint]{aastex}

\slugcomment{Accepted for publication in ApJ}

\shorttitle{Intermediate-Redshift X-ray Groups}
\shortauthors{Mulchaey et al.}

\begin{document}


\title{X-ray Selected Intermediate-Redshift Groups of Galaxies}


\author{John S. Mulchaey}
\affil{The Observatories of the Carnegie Institution of Washington, 813 Santa Barbara St., Pasadena, 91101}
\email{mulchaey@ociw.edu}

\author{Lori M. Lubin, Chris Fassnacht}
\affil{Department of Physics, University of California at Davis, 1 Shields Avenue, Davis, CA 95616}

\author{Piero Rosati}
\affil{European Southern Observatory, Karl-Schwarzschild-Strasse 2, D-85748 Garching, Germany}

\and 

\author{Tesla E. Jeltema}
\affil{The Observatories of the Carnegie Institution of Washington, 813 Santa Barbara St., Pasadena, 91101}


\begin{abstract}
We present spectroscopic confirmation of nine moderate redshift galaxy
groups and poor clusters selected from the \textit{ROSAT} Deep Cluster
Survey. The groups span the redshift range z $\sim$ 0.23 --0.59 and
have between 4 and 20 confirmed members. The velocity dispersions of
these groups range from $\sim$ 125 to 650 km s$^{-1}$. Similar to
X-ray groups at low redshift, these systems contain a significant
number of early-type galaxies. Therefore, the trend for X-ray luminous
groups to have high early-type fractions is already in place by at
least z $\sim$ 0.5.  In four of the nine groups, the X-ray emission is
clearly peaked on the most luminous early-type galaxy in the group.
However, in several cases the central galaxy is composed of multiple
luminous nuclei, suggesting that the brightest group galaxy may still be
undergoing major mergers.  In at least three (and possibly five) of
the groups in our sample, a dominant early-type galaxy is not found at
the center of the group potential.  This suggests that many of our
groups are not dynamically evolved despite their high X-ray
luminosities.  While similar systems have been identified at low
redshift, the X-ray luminosities of the intermediate redshift examples
are one to three orders of magnitude higher than those of their low
redshift counterparts. We suggest that this may be evidence for group
downsizing: while massive groups are still in the process of
collapsing and virializing at intermediate redshifts, only low-mass
groups are in the process of forming at the present day.

\end{abstract}


\keywords{galaxies: clusters: general --- galaxies: elliptical and lenticular, cD ---
galaxies: evolution}


\section{Introduction}

Groups of galaxies constitute the most common galaxy associations,
containing as many as 50--70\% of all galaxies \citep{tur76,gel83,eke06}.
They are, therefore, an important laboratory for studying the
processes associated with galaxy formation and evolution.  In recent
years, optical and X-ray studies of groups at low redshift have
provided new insights into these important systems. In particular,
there are strong correlations between the morphological composition of
the luminous galaxies, the velocity dispersion, and the presence of
X-ray emission \citep{zab98,mul98, mul03,osm04}.  Specifically,
diffuse X-ray emission is found almost exclusively in those groups
dominated by early-type galaxies.  In turn, the early-type fraction is
strongly correlated with the group velocity dispersion and, thus, the
group mass.

In the most luminous X-ray groups, the brightest group galaxy (BGG) is
always a very massive elliptical, located at the peak of the X-ray
emission \citep{ebe94,mul96, mul98,hel00,mul03,osm04}.  As the peak of
the X-ray emission is likely coincident with the center of the group,
this implies that the BGG lies at the center of the group
potential. Indeed, the position of the BGG is also indistinguishable
from the center of the group potential as defined by the mean velocity
and projected spatial distribution of the galaxies \citep{zab98}. The
fact that the BGG is located at the center of the potential suggests
the formation of the BGG is intimately linked to the formation and
evolution of the group itself.

Given their relatively low velocity dispersions, groups of galaxies
provide ideal sites for galaxy-galaxy mergers
\citep{bar85,aar80,mer85,mil04,tay05,tem06}.  
This implies that significant changes in
the star formation rates and morphological appearance of galaxies may
be occurring in groups. To better understand how galaxies evolve in
the group environment, groups must be observed over a wide range of
cosmic time.  However, observations of groups at even moderate
redshifts have been limited because of the difficulty of finding
groups given their low galaxy densities.  \citet{all93}
photometrically selected a sample of groups at intermediate redshifts
by targeting known radio galaxies. Their study suggested a progressive
bluing in the galaxy population. Small samples of groups at higher
redshifts have also been found in deep redshift surveys
\citep{lub98,coh00} and around lensed quasars
\citep{rus01,fas02,nai02,ray03,gra04,fau04,wil06}.  The recent
completion of very large redshifts surveys now allow large group
samples to be kinematically-defined from moderate redshifts up to z
$\sim$ 1 \citep{car01,ger05}.  Wilman et al. (2005a,b) studied a large
sample of groups at moderate redshifts selected from the CNOC2 survey
and found that the fraction of group members undergoing significant
star formation increases strongly with redshift out to z $\sim$ 0.5.
Therefore, there is evidence for some evolution in the group
environment in the last $\sim$ 5 billion years at least among
optically-selected group samples.

X-ray emission from the hot intragroup medium provides another way to
identify candidate groups at high redshifts. 
\textit{ROSAT} was the first X-ray telescope capable of finding such systems
and large numbers of group candidates at intermediate redshifts were found
in deep surveys with this telescope 
\citep{ros95,sca97,bur97,ros98,jon98,vik98,rom00,ada00,per02,jon02,bur03}.
The ROSAT surveys suggest there is little or no 
evolution of the X-ray luminosity
function of groups and poor clusters out to at least z=0.5.
More recently, \citet{wil05} arrived at a
similar conclusion using twelve groups and clusters from the early
data taken as part of the \textit{XMM} Large-Scale Structure (LSS)
Survey.  Upon completion, the XMM-LSS survey will provide a large
sample of X-ray selected groups and poor clusters out to redshifts of
z $\sim$ 0.6 or higher.

In this paper, we provide the first results from an extensive
multi-wavelength study of nine X-ray selected galaxy groups and poor
clusters in the redshift range 0.2 $<$ z $<$ 0.6 selected from deep
\textit{ROSAT PSPC} pointings.  Our data allow the first detailed look
at the morphological composition of X-ray groups at intermediate
redshifts.  A detailed study of the X-ray properties of six of these
systems based on \textit{XMM-Newton} observations is provided in a
companion paper (Jeltema et al. 2006; hereafter Paper II). We assume a
$\Lambda$ cold dark matter cosmology with $\Omega$$_{m}$=0.27,
$\Lambda$=0.73, and H$_{\rm o}$=70 km s$^{-1}$ Mpc$^{-1}$ throughout
this paper.

\section{Sample Selection}

We select our intermediate redshift group candidates from the
\textit{ROSAT} Deep Cluster Survey (RDCS; \citep{ros98}).  
As the deepest of the ROSAT pointing surveys, the RDCS is the best
suited for selecting low luminosity systems (i.e. groups) at intermediate
redshifts.
The RDCS
used a wavelet-based technique to search for extended sources in deep
\textit{ROSAT PSPC} pointings.  A full description of the X-ray
analysis and object selection technique is provided in \citet{ros95}.
To identify the poorest galaxy systems in the RDCS, we restrict our
sample to objects with X-ray luminosities between $\sim$ 2 $\times$
$10^{42}~h_{70}^{-2}~{\rm ergs~s^{-1}}$ and $\sim$ 2 $\times$
10$^{43}~h_{70}^{-2}~{\rm ergs~s^{-1}}$ in the 0.5--2 keV band. This
corresponds to the range of nearby X-ray luminous groups
\citep{mul03,osm04}. As we are interested in moderate redshift groups, 
we further restrict our
sample to objects at z $>$ 0.2.  Figure 1 shows the distribution of redshifts and
X-ray luminosities for RDCS groups as defined above. The majority of the 
systems are at the low redshift end (0.2 $<$ z $<$ 0.3) as is expected from a 
flux-limited X-ray survey. However, for our initial study we have concentrated 
mostly
on the z $>$ 0.3 groups, as these provide the longest time baseline 
for comparison with nearby groups. 
Furthermore, we have obtained \textit{XMM-Newton} data primarily for the 
highest luminosity systems, as these targets tend to require the shortest 
exposure times.


Figure 2 shows contours of the diffuse X-ray emission overlaid on
optical images of the six fields for which we have \textit{XMM-Newton}
data (Paper II).  The three systems with only \textit{ROSAT} data are
shown in Figure 3.  The association of the X-ray emission with an
overdensity of galaxies is apparent in many cases.

\section{Observations}

\subsection{Spectroscopic Data}

To determine group membership, we obtained optical spectra for
galaxies in each field using multi-object spectrographs at the
Palomar, Las Campanas and Keck Observatories. Spectroscopic candidates
for each group were selected from the original R or I band images
taken for the RDCS with the KPNO and CTIO 4m telescopes. 
The program SExtractor \citep{Ber96} was used to classify objects as
stars or galaxies and to measure magnitudes. For the present work, we
considered all objects with a \lq\lq stellarity-index\rq\rq \ of less
than 0.5 as galaxies. At the redshifts of these groups, the typical
seeing (0.8--1.5$''$) corresponds to $\sim$ 3--10 kpc. Thus, we are
unable to cleanly distinguish galaxies smaller than this size from stars.
This has some impact on the selection of objects for our
spectroscopic survey, although the effect appears to fairly small
(based on the spectroscopy less than 10\% of the targeted objects are stars).
Priority was given to the brightest objects in
each field. No color information was used to select spectroscopic
candidates. This selection method has the advantage that it avoids 
biasing our study towards certain types of group galaxies (i.e. 
red galaxies), but comes at the price of a higher fraction of the 
spectroscopic targets being non-group members.
Typically, three multi-slit masks were created for each
group, with each mask containing between 15 and 20 objects.

All of the Palomar spectra were taken with the COSMIC spectrograph
\citep{kel99} on the 5.1m Hale telescope between 1999 and 2001. The
instrument configuration resulted in spectra covering the wavelength
range $\sim$ 3500--9800 \AA \ with a spectral scale of $\sim$ 3.1
\AA/pixel.  The two southern targets were observed with the WFCCD
spectrograph on the du Pont 100-inch telescope at Las Campanas
Observatory, Chile in October, 2000. The spectra cover the wavelength
range $\sim$ 3800--9500 \AA \ with a spectral scale of $\sim$ 4
\AA/pixel. A similar observing scheme was used for the COSMIC and
WFCCD observations. Typically, two one-hour integrations were taken
for each mask, with an arc and flatfield exposure taken at the
completion of each mask exposure.  A CuAr arc lamp was used for the
COSMIC data and a HeNeAr lamp was used with the WFCCD.  Finally, one
group (RXJ0329.0+0256) was observed with LRIS \citep{okelris} on Keck
I during December 2004.  The LRIS spectra are centered at $\sim$ 6000
\AA \ and have a total wavelength range of $\sim$ 2500 \AA.

The COSMIC and WFCCD data were reduced using IRAF. First, the overscan
regions of the CCD chips were used to estimate and subtract the bias
from each frame. Flatfield exposures were then used to construct a
normalized flatfield frame by performing a low-order fit in the
dispersion direction. The science frames were then divided by this
normalized flatfield to correct for the pixel-to-pixel response of the
detector. A distortion correction was applied to each frame to align
the spectra along the rows of the detector. The positions of the
objects on the slit and corresponding sky regions were then defined
interactively using the IRAF package Apextract.  The spectra were then
sky-subtracted and extracted to produce one-dimensional
spectra. Wavelength calibrations were determined for each spectrum
from the arc exposures.

The LRIS spectroscopic data were reduced with custom scripts written
to process multislit observations.  The scripts serve as front ends to
standard {\sc iraf} tasks.  The processing included overscan
subtraction, flat-field correction, cosmic-ray rejection, and sky
subtraction.  There were three science exposures obtained through each
slitmask, with small dithers in the spatial direction made between
exposures.  The background-subtracted science exposures for each
slitmask were co-added, and then one-dimensional spectra were
extracted from the combined two-dimensional files.  The wavelength
solutions were determined from arclamp exposures that were obtained
immediately after the first science exposure for each slitmask.

Redshifts were measured using an IRAF script called redsplot
(T. Small, private communication). This task allows the user to
interactively identify a potential line feature in the spectrum and
then plot the locations of other line features from a spectral line
list. Redshifts are then measured from the centroid of each line. We
adopt the average measurement from all of the line features as the
final redshift for each object. The standard deviation of all of the
measurements is used to estimate the error. The redshifts for all of
the group members in this paper were based on a minimum of three
distinct line features. More typically, five or six features were used
for the redshift determination. In total, we measure redshifts for 169
galaxies. This corresponds to a $\sim$ 65\% success rate. A significant 
fraction of the objects with measured redshifts have 
emission lines in their spectra ($\sim$ 46\%). However,
the emission-line fraction among
confirmed 
group members is much lower ($\sim$ 20\%). This result is not surprising
given the large population of early-type galaxies that we find in these 
groups (\S 4.3). Many of the group members (63\%) 
with emission lines have spectra 
consistent with the presence of an AGN. A detailed discussion of the emission-line
properties of the group galaxies is deferred to a future paper.

\subsection{Imaging Data}

During the course of the Palomar and Las Campanas observing runs
described above, images of each group were taken when conditions were
photometric. The images were taken with a Kron-Cousins R filter from
the Harris set. Typically, we took a series of nine five-minute
exposures for each field, resulting in a total integration time of 45
minutes. The images were reduced using standard techniques in
IRAF. The bias level was determined from the overscan region of the
CCD and subtracted from the images. Flat-fielding was accomplished
using dome flats.  The images were flux-calibrated using observations
of standard star fields from \citet{lan92}. SExtractor \citep{Ber96} was used to
measure a magnitude for each galaxy.
We adopt the \lq\lq MAG\_AUTO\rq\rq \ option 
for our total magnitudes. 

In addition to providing magnitudes, the R-band data were also used to
determine the morphologies of the group members. Given the distances
of these objects and the quality of these images 
it is generally not possible to make detailed morphological
classifications with the groundbased data. For this reason, we
restrict our classifications to \lq\lq early\rq\rq \ and \lq\lq
late\rq\rq \ type galaxies.  Each group member was visually classified
by J.S.M. into one of the two types. Higher resolution \textit{Hubble
Space Telescope} images taken with the \textit{WFPC2} are available
for all nine fields. However, given the small field of view of
\textit{WFPC2}, only about half of the spectroscopically confirmed
group members were observed by \textit{HST}. Independent
classifications were made for each group galaxy observed with
\textit{HST}. In general, the agreement between the groundbased
classifications and the \textit{HST}-based classifications are good.
There is a disagreement in the type in $\sim$ 13\% of the galaxies. In
all of these cases, the \textit{HST} data suggest that a late-type
galaxy has been misidentified as an early-type object. This is not too
surprising because the \textit{HST} images have the ability to reveal
faint spiral structure that is not apparent in the lower resolution
ground-based images.

Table 1 lists the J2000 coordinates, R-band magnitude, redshift,
redshift error and type of redshift measurement (abs=absorption,
em=emission and abs + em = combination of absorption and emission
lines) for each galaxy in our survey with a measured redshift.

\section{Results and Discussion}

\subsection{Group Membership}

In most cases, the identification of the group in redshift space is
trivial as the spatial distribution of galaxies on
the sky coincide with the X-ray emission.  However, for a few of these
fields, there are several different galaxy systems superposed along
the line of sight.  This leads to some ambiguity about the true
redshift of the X-ray system in two of the nine systems studied
here. The RXJ1205.9+4429 field contains two
significant galaxy systems, one at z $\sim$ 0.35 and another z $\sim$
0.59. The \textit{XMM-Newton} observation of this field shows that the
X-ray emission is clearly centered on a luminous early-type galaxy
that is part of the z=0.59 system (Paper II). Thus, we adopt this
value as the redshift of this group. We note that the preliminary RDCS
redshift corresponded to the lower redshift system. Therefore, the
X-ray luminosity is actually considerably higher than originally
reported in the RDCS. The true X-ray luminosity of this system is high
enough that it does not meet our original selection criterion,
suggesting this is likely a much richer system than the other objects
in our sample. \citet{ulm05} have recently published a detailed study
of the X-ray and optical properties of this system and conclude that
it is a fossil group. However, our spectroscopy and imaging data
indicate that the magnitude difference between the brightest and
second brightest confirmed member is $\sim$ 1.2 mag. in the
R-band. Thus, this system is not a fossil group by the standard
definition usually adopted in the literature \citep{jon03}.

 In the case of RJX0341-4459 field, we measure redshifts for five
galaxies in a system at z $\sim$ 0.41. The five galaxies have a
spatial distribution similar to the X-ray emission. However, there are
also three foreground galaxies distributed over the same area. As
these three objects have very different redshifts, they are not part
of a single galaxy system. Thus, we believe the X-ray emission is most
likely associated with the system at z $\sim$ 0.41, although higher
resolution X-ray images are required to be sure. Examples like this
demonstrate the difficulty sometimes encountered when trying to
identify low galaxy density systems (i.e. groups) at high redshift
even when X-ray emission is present.

We determine group membership for each system using the ROSTAT package
\citep{bee90}. We start by considering all galaxies within $\pm$3000
km s$^{-1}$ of the group's mean velocity. This is a large enough range
to include all potential group members. We then calculate the biweight
estimators of location (mean velocity) and scale (velocity
dispersion). Objects with velocities greater than three times
$\sigma$$_{\rm biwt}$ are then removed from the sample and a new mean
location and scale are calculated. This process is repeated until
there are no more objects to be clipped. This procedure resulted in
the removal of one galaxy from three of the groups and none from the
remainder. Figure 4 shows the velocity distributions of each member 
relative to the final mean velocity of the group.
The final mean velocity and velocity dispersion are given
in Table 2. For all of the systems studied here, the classical
velocity dispersion (i.e.  $\sigma_{\rm Gauss}$, the Gaussian
estimator) is in good agreement with the biweight velocity dispersion
estimate. For approximately half of our sample, the velocity
dispersions are based on only $\sim$ 5 velocity measurements. These
velocity dispersions are rather uncertain. Studies of low redshift
X-ray groups suggest velocity dispersions based on a small number of
galaxies can be significantly underestimated \citep{zab98}.

\subsection{The L$_{\rm X}$-$\sigma$ Relationship}

Figure 5 shows the L$_{\rm X}$-$\sigma$ relationship for our nine
groups along with the sample of nearby groups from \citet{osm04} and
the moderate redshift X-ray groups from \citet{wil05}. As can be seen
from the figure, several of the groups fall significantly off the
relationships found for nearby groups and clusters.  The two most
deviant points in our sample correspond to the RXJ1334.0+3750 and
RXJ1648.7+6019 groups.  These are the two groups from our
\textit{XMM-Newton} survey where the X-ray emission is not centered on
an early-type BGG (Paper II).  In both cases, these groups have very
low velocity dispersions for their given X-ray luminosities.  There
are several possible explanations for why these groups fall so far off
the relationships found for low redshift groups and clusters.  First,
our velocity dispersions estimates for these groups may be
artificially low because they are based on relatively small numbers (6
and 8 members, respectively).  \citet{zab98} find that velocity
dispersions estimated from the five brightest galaxies can be
underestimated by as much as a factor of three. A similar factor would
bring our two most deviant groups in agreement with the relationship
found for nearby groups and clusters. This idea can be tested by
obtaining more velocity measurements.  Second, the X-ray luminosities
of these groups may have been enhanced or contaminated in some
way. For example, the observed X-ray emission may be dominated 
by galaxy emission that is unresolved with our \textit{XMM-Newton}
observations. While we believe this is very unlikely given the high X-ray
luminosities of our groups and the extent and morphology of the X-ray
gas, we cannot rule this possibility out without higher resolution
X-ray images.
 Thirdly, the velocity dispersions may have been reduced
in some way. \citet{hel05} have studied several nearby groups that
fall off the L$_{\rm X}$-$\sigma$ relationship in a similar manner to
our groups (although the X-ray luminosities of the nearby groups are
nearly two orders of magnitude lower than the present examples). They
suggest several physical mechanisms that could reduce the velocity
dispersions including dynamical friction and tidal heating. They also
suggest that orientation effects can lead to an artificially low
observed velocity dispersion.  Finally, the low velocity dispersions
could be an indication that these groups are in the process of
collapsing for the first time and therefore the measured velocity
dispersions do not yet accurately reflect the depth of the group
potential (see Section 4.5).

\subsection{Morphological Composition}

Studies of X-ray groups at low redshift have revealed a very strong
tendency for these systems to be dominated by early-type galaxies
\citep{ebe94,pil95,hen95,mul96,zab98}. Table 2 lists the early-type
fraction for our groups (based on the \textit{HST} morphological
classifications, where possible).  For all but one of our objects, the
early-type fractions are in the range $\sim$ 0.4--0.8. For the four
groups with just four to six members known, these fractions could be
somewhat over-estimated as our analysis is restricted to the brightest
group members, which tend to be ellipticals \citep{zab98}. However,
even for the groups with many more members identified, the early-type
fractions are comparable to those of rich clusters sampled out to
similar radii \citep{whi93}. Thus, the trend for X-ray groups to
contain a large number of early-type galaxies appears to be in place
out to at least z $\sim$ 0.5. The one exception in our sample is the
RXJ0210.4-3929 group, which based on the HST imaging
is dominated by spiral galaxies. The large number of spirals in this 
system
make it very unusual among X-ray groups at both low and moderate
redshifts.

A correlation between early-type fraction and velocity dispersion has
been noted for nearby group samples \citep{hic88,zab98,osm04},
suggesting that galaxy morphology is related to the depth of the group
potential. For groups with well-determined membership, the
relationship is surprisingly robust \citep{zab98}. In Figure 6, we
plot these quantities for our sample along with the low redshift data
from \citet{zab98}. Among our moderate redshift groups, there is
considerable scatter and no indication of a trend between early-type
fraction and velocity dispersion. We suspect much of this scatter is
an indication that neither quantity is well-determined for most of our
groups. However, we note that the two groups in our sample with
membership data comparable to that of the \citet{zab98} sample
(i.e. $\sim$ 20 known members) do appear to follow the trend found at
low redshift. In fact, these two groups suggest that the relationship
found by \citet{zab98} extends to the range of poor clusters. As noted
by \citet{zab98}, the relationship cannot be the same as for rich
clusters as it would predict an unphysical early-type fraction for
clusters with velocity dispersions above $\sim$ 800 km s$^{-1}$.

\subsection{The Brightest Group Galaxy}
 
Previous work on low redshift X-ray groups indicates that the X-ray
emission is usually centered on a luminous elliptical galaxy
\citep{ebe94,mul98,hel00,mul03,osm04}.  In almost every case, this
elliptical is the most luminous galaxy in the group.  As the peak in
the X-ray emission is likely coincident with the center of the group
potential, this implies that the brightest group galaxy (BGG) lies at
the center of the potential.

Unfortunately, it is difficult to define the peak of the X-ray
emission for the present sample from the low signal-to-noise
\textit{ROSAT} images.  However, six of the nine groups have now been
observed by \textit{XMM-Newton} and four of these are consistent with
the X-ray peak being coincident with the brightest group elliptical
(Paper II).  In all four groups with a central BGG, the radial
velocity of the BGG is consistent with the mean velocity of the group
within the velocity errors. Thus, similar to the case found for nearby
X-ray groups \citep{zab98}, the BGG is likely at or near the center of
the group potential in these systems.

However, for three of the four groups where we find a dominant BGG,
the central object appears to be composed of multiple nuclei (see
Figure 7).  In the two most spectacular cases (RXJ0720.8+7109 and
RXJ1256.0+2556), the central object has three components.  Although
multiple nuclei in CD galaxies in clusters are fairly common
\citep{hoessel80,schneider83,lauer88}, in the majority of cases there
is a large magnitude difference between the various components. In
contrast, for both of our three nuclei systems, the second 
nuclei have R-band magnitudes within $\sim$ 0.5 mag. of the brightest
component.  In the case of the RXJ0720.8+7109 group, the second nucleus
is the second brightest galaxy in the group.  Previous studies of
multiple nuclei in clusters have found large velocity offsets between
the various components in some cases, indicating these systems are not
bound and are thus not in the process of merging
\citep{merritt84,Tonry85,smith85}.  For the RXJ0720.8+7109 group, we
obtained a spectrum of the two brightest components and find a radial
velocity difference of $\sim$ 200 km s$^{-1}$. Given the typical
errors on our velocity measurements ($\sim$ 100 km s$^{-1}$), our data
are consistent with a similar radial velocity for the two components.
Thus, the two components may be bound. For the other triple system,
RXJ1256.0+2556, we only obtained a velocity for the central component,
so we cannot infer anything further about the nature of the multiple
nuclei.

As noted above, in only four of the nine groups in our sample is the
X-ray emission clearly centered on an early-type galaxy. In two of the
other groups the most luminous galaxy is an elliptical, but the
existing X-ray data are not sufficient to determine an X-ray center
(RXJ0341.9-4459 and RXJ1347.9+0752).  Thus, we cannot draw strong
conclusions for these two groups as to whether the X-ray emission is
peaked on the dominant elliptical galaxy or not.  For both groups, the
brightest elliptical is offset in velocity from the mean velocity of
the group by several hundred kilometers per second. 
However, we have very few velocity measurements for
both systems, so the mean velocity of the group is not well-determined
and we cannot draw strong conclusions regarding a
potential offset of the BGG from the group center. 

However, the remaining three groups in our sample deviate strongly
from the low redshift trend for there to be a dominant elliptical
galaxy at the center of the group potential.  The RXJ1334.0+3750 group
contains a dominant elliptical, but the peak of the X-ray emission is
offset significantly from this galaxy (Paper II).  The velocity of
this galaxy is consistent with the mean velocity of the group within
the errors on each measurement. However, given the very low velocity
dispersion of this system ($\sigma$=121$^{+58}_{-45}$) and the small
number of known members (6), we cannot draw strong conclusions with
the existing velocity data.

The most luminous galaxy in the RXJ1648.7+6019 group is also an
elliptical, although the group contains several galaxies of 
comparable luminosity.  Furthermore, the \textit{XMM-Newton} data
suggest the X-ray emission is not centered on any particular galaxy,
but is instead distributed in a chain morphology similar to the
distribution of galaxies near the group center (Paper II). The
brightest elliptical is also offset in velocity from the mean velocity
of the group by more than 200 km s$^{-1}$. This provides further
evidence that this galaxy is not at the center of the group potential.

A chain-like morphology is also found in the RXJ0210.4-3929 group. In
this case, the brightest galaxies in the chain are
spirals. Unfortunately, we do not have \textit{XMM-Newton} data for
this system, so we cannot be sure where the X-ray emission peaks.
Regardless, the most luminous early-type galaxy near the group center
is nearly a magnitude fainter than the brightest spirals and has a
velocity offset by nearly 400 km s$^{-1}$ from the mean velocity of
the group. Thus, this group also lacks a dominant central early-type
galaxy.

\subsection{Evidence for Group Downsizing}

As discussed in the last section, at least three (and potentially
five) of the nine groups in the present sample do not appear to have a
central dominant early-type galaxy.  Furthermore, in three of the four
groups where the X-ray emission is centered on a BGG, the central
galaxy is not a single object, but rather is composed of multiple
components. These observations suggest that most of the groups in our
sample are not dynamically evolved. Instead, we appear to be catching
them in the process of virialization.
 
The global X-ray properties of these groups are consistent with the
properties of more dynamically relaxed systems, however (Paper II).
This suggests that the X-ray properties of groups are already in place
early in the formation of these systems.  Specifically, the intragroup
medium properties appear to be largely set prior to the BGG
experiencing its last major merger and settling at the center of the
group potential.  This scenario would also explain the lack of
evolution observed in the X-ray luminosity function of groups out to z
$\sim$ 0.5 \citep{ros95,jon02,wil05} despite the morphological peculiarities
we find over the same redshift interval.  If true, the temperature of
the hot gas component may provide a better indication of the global
group potential early on than the velocity dispersion of the galaxies.
This might explain the very low velocity dispersions observed for the
RXJ1334.0+3750 and RXJ1648.7+6019 groups: The X-ray temperatures of
these systems reflect the massive group potentials, but the velocity
dispersions of the galaxies do not yet accurately probe the group
mass. Cosmological simulations of groups that include both the
intragroup gas and galaxies may be able to test this idea.

The late assembly of the BGG in groups is consistent with the results
of simulations in hierarchical cosmological models \citep{dub98}. The
groups in our sample appear to cover a range in dynamical state and
can therefore provide some clues into the formation process of the
BGG. The RXJ0210.4-3929 and RXJ1648.7+6019 groups do not yet contain a
dominant early-type galaxy and thus they are likely at the earliest
stages of group formation. The morphological compositions of these
groups are very different, with RXJ0210.4-3929 consisting mostly of
spirals and RXJ1648.7+6019 mostly of early-type galaxies. This
suggests that both early and late type galaxies can be the dominant
contributor to the final merger product. The groups with a multiple
component BGG are likely much further along in the virialization
process.  In fact, the BGG in these groups is probably undergoing its
final major merger. Finally, only one of the groups in our sample is
consistent with being a relaxed, virialized system
(RXJ0329.0+0256). In this case, the BGG is at the center of the group
potential as determined from both the X-ray emission and the velocity
distribution of the group members and is unlikely to undergo any more
major mergers.

The fact that many of our intermediate redshift groups do not have
a dominant central elliptical is somewhat surprising given that such
groups appear to be very rare among local X-ray group samples
\citep{mul03,osm04}.  One potential concern in comparing our
intermediate redshift groups to local samples is that the best studied
local samples were not selected in a similar manner. In fact, the
largest ROSAT surveys of groups were performed with very heterogeneous
samples of groups mostly drawn from optical catalogs
\citep{mul00,mah00,hel00,mul03,osm04}. To allow a better comparison to
low redshift systems, we have selected a sample of nearby X-ray groups
and poor clusters from two surveys based on the \textit{ROSAT} All-Sky
Survey: the NORAS \citep{boh00} and REFLEX \citep{boh04} group and
cluster samples.  From each survey, we have selected all of groups and
clusters with X-ray luminosities between $\sim$ 2 $\times$
$10^{42}~h_{70}^{-2}~{\rm ergs~s^{-1}}$ and $\sim$ 2 $\times$
10$^{43}~h_{70}^{-2}~{\rm ergs~s^{-1}}$ in the \textit{ROSAT} band
(i.e. the corresponding selection criterion for our intermediate
redshift sample) with z $\le$ 0.05. Eliminating duplicate entries from
the two catalogs produces a sample of 74 X-ray luminous groups and
clusters. Unfortunately, the vast majority of these systems have not
been previously studied in detail in either the optical or X-ray
bandpasses. However, a literature search reveals that 19 of the 74
systems have deeper X-ray images published. The existing data for this
subset of groups suggests these X-ray selected systems follow the
trends found among the more heterogeneously selected nearby group
samples. In particular, in all 19 of the groups, the X-ray emission is
centered on a luminous early-type galaxy. Furthermore, we find no
multiple-nuclei examples among the 19 nearby BGGs.  This suggest that the
differences we find between our intermediate redshift systems and
local samples are not the result of a selection effect. Rather, the
intermediate redshift groups appear to be less dynamically evolved
than present day luminous X-ray groups.

A closer examination of low redshift samples suggests that there are
some local examples of X-ray groups without a central BGG
\citep{mul03,osm04,ras06}. However, the X-ray luminosities of these
systems are one to three orders of magnitude lower than the X-ray
luminosities of our moderate redshift groups.  Among the $\sim$ 60 low
redshift X-ray groups that have been studied in detail with
\textit{ROSAT}, the most luminous example of a system without a
early-type BGG is the NGC 5171 group (L$_{\rm X}$ $\sim$ 3 $\times$
10$^{42}$ erg/s; \citet{osm04}).  Thus, among the most X-ray luminous
(L$_{\rm X}$ $>$ 5 $\times$ 10$^{42}$ erg s$^{-1}$) groups in the
nearby universe, there appear to be no known counterparts to the
systems we find at intermediate redshifts.  The failure to find 
nearby examples of such systems suggests that the most X-ray luminous
groups have largely reached virialization by z $\sim$ 0. This suggest
that we are witnessing group downsizing: While the most luminous (and
thus most massive) groups are still in the process of virializing at
intermediate redshifts, this process is restricted to much less
luminous (and thus less massive) systems at the present day.

\section{Summary}

We have performed multi-object spectroscopy in the fields of nine
candidate X-ray groups selected from the \textit{ROSAT} Deep Cluster
Survey. The velocity dispersions derived from our data span the range
expected for groups and poor clusters. We have used groundbased and
HST images of these fields to quantify the morphological compositions
of these systems. We find that like low redshift X-ray groups, these
systems contain a substantial population of early-type galaxies.
Therefore, the large early-type fractions in X-ray groups are in place
by at least z $\sim$ 0.5.

In four of our nine groups, the X-ray emission is centered on a
dominant early-type galaxy. In these cases, the velocity of the
dominant galaxy is consistent with the mean velocity of the group,
suggesting these galaxies are at the center of the group potential.
However, in three of these four groups, the central galaxy is composed
of multiple components, suggesting the BGG is still undergoing major
mergers. This idea can be tested with more detailed spectroscopy of
the multiple nuclei.

In at least three and potentially five of our groups, we find no
evidence for a dominant central early-type galaxy. In several cases, a
dominant elliptical exists, but it is not at the center of the group
potential as determined by the X-ray emission or velocity
distribution.  In addition, two of the groups in the present sample do
not contain a dominant early-type galaxy at all.

The fact that a large fraction of our intermediate redshift groups
contain a BGG with multiple components or contain no central BGG at all
suggests that these systems are not dynamically evolved.  However, the
X-ray properties of these systems are similar to those of nearby
virialized groups, suggesting that the X-ray emission in groups likely
reflects the global properties of the potential earlier than the
velocity distribution of the member galaxies.  A comparison of our
moderate redshift sample with similarly selected groups at low
redshift indicates that the most X-ray luminous groups have reached a
virialized state by the present time. However, there are examples of
lower X-ray luminosity systems at low redshift that do not contain a
central BGG. This may be evidence for group downsizing: While massive
groups were still in the process of collapsing and virializing at
intermediate redshifts, only lower mass systems are forming at the
present time.

While the current study has uncovered some interesting results, much
better data will be required to confirm our conclusions.  We are in
the process of obtaining higher quality spectroscopy for these systems
which will allow a much more detailed analysis of their dynamical
state and a first look at the properties of the individual galaxies
in these groups.


\acknowledgments

We thank Roy Gal for advice on reducing the COSMIC data. We also
acknowledge useful discussions with Alan Dressler, Mike Gladders,
Daisuke Kawata and Dan Kelson. We also thank Somak Raychaudhury for
suggestions that significantly improved this paper and Trevor Ponman 
for providing us with the fits to the GEMS data prior to publication.
JSM acknowledges support from NASA
grants NNG04GC846 and NNG04GG536 and HST grant G0-08131.01-97A.



{\it Facilities:} \facility{ROSAT},\facility{Du Pont}, \facility{Hale}, \facility{Keck}, 
\facility{HST}


\vfill\eject

\includegraphics[width=5in]{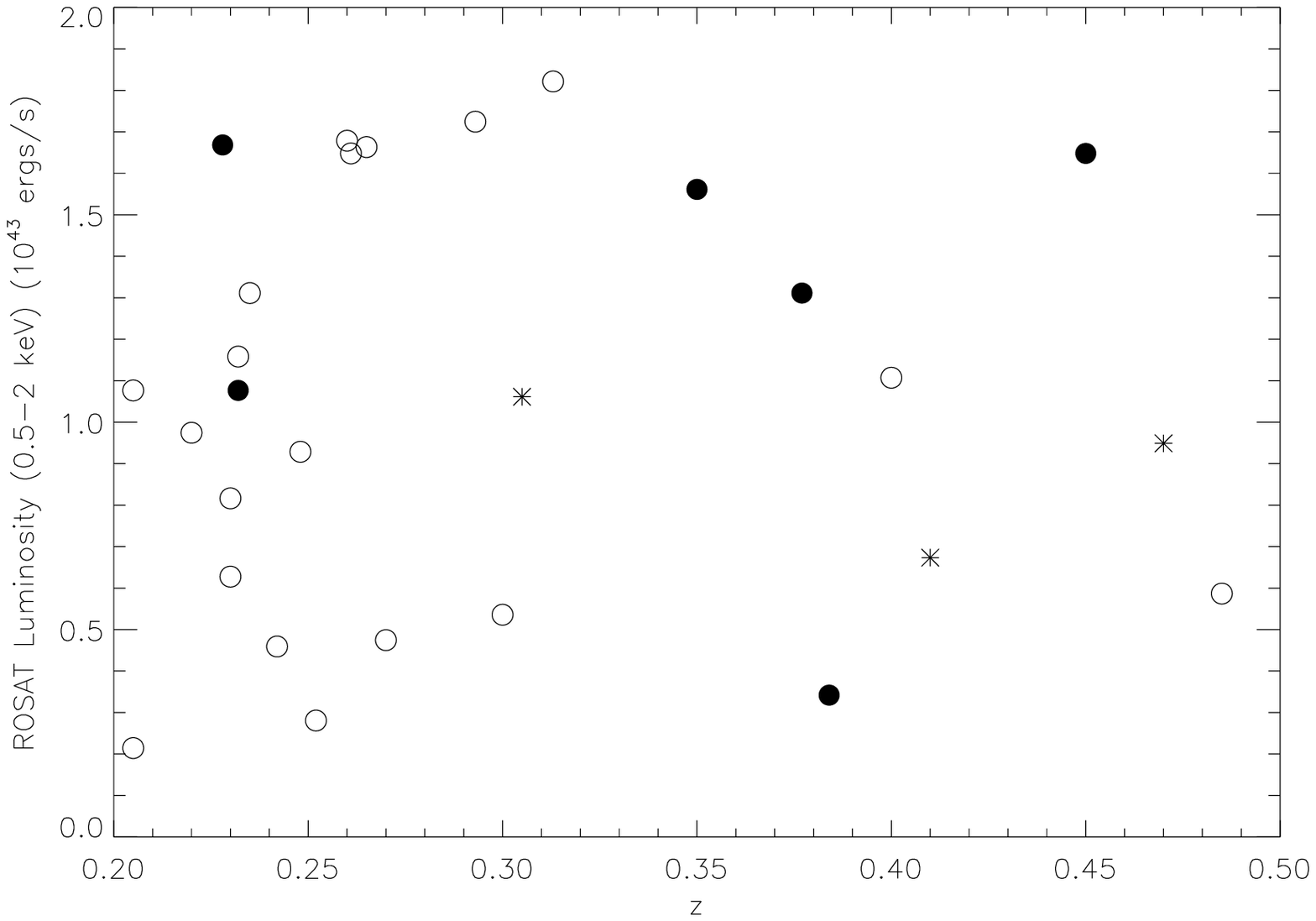}
\vskip 0.2cm
\noindent Figure 1 - Relationship between redshift
 and X-ray luminosity (L$_{\rm X}$) for the groups in the RDCS 
 sample. Objects in the subsample discussed in this paper with
\textit{XMM-Newton} observations are shown as filled circles, while the
 groups in the present paper with only \textit{ROSAT} observations are shown as stars.
The redshifts and X-ray luminosities plotted are taken from the original RDCS survey.

\vfill\eject

\includegraphics[width=2.5in]{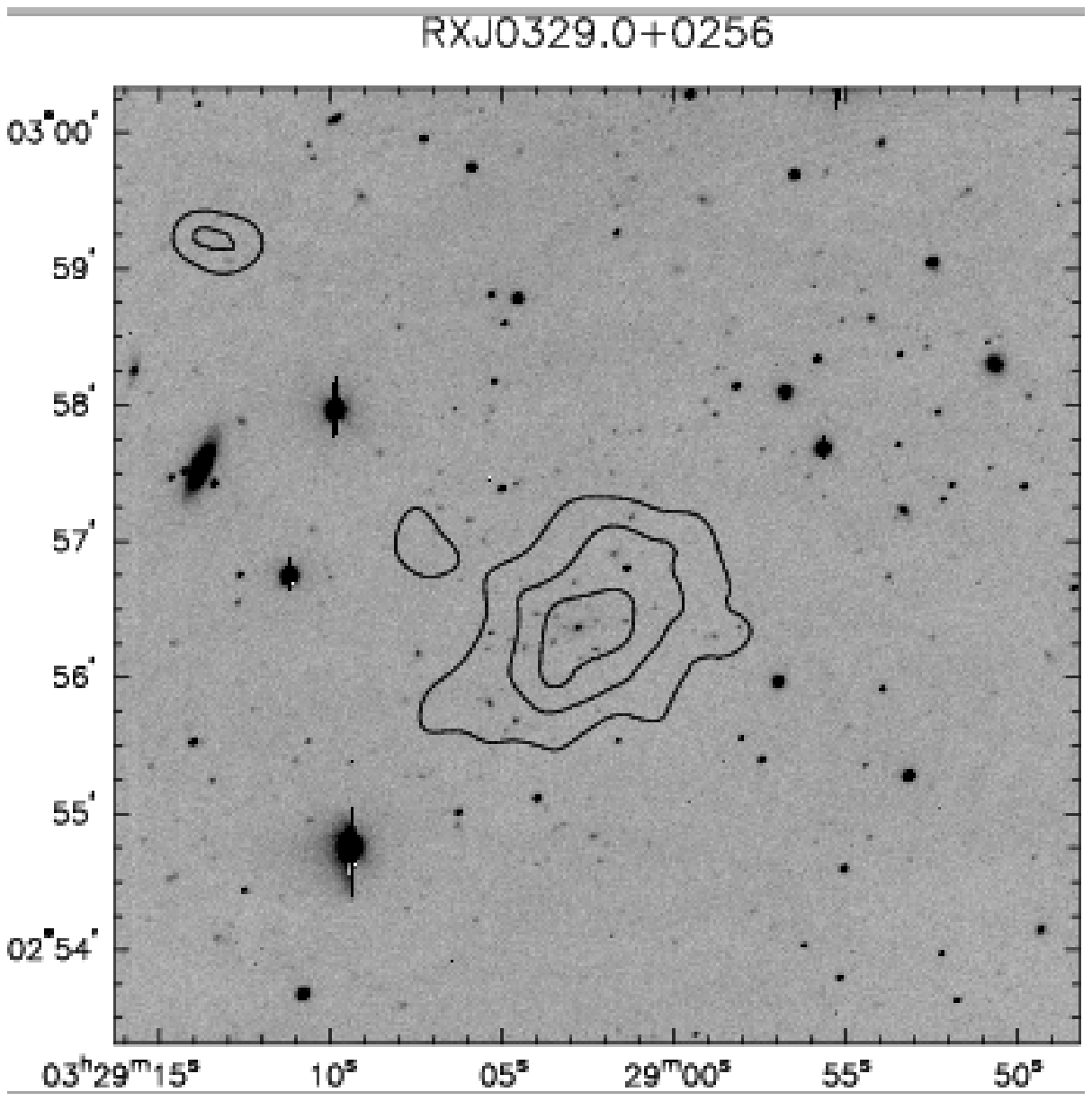}
\includegraphics[width=2.5in]{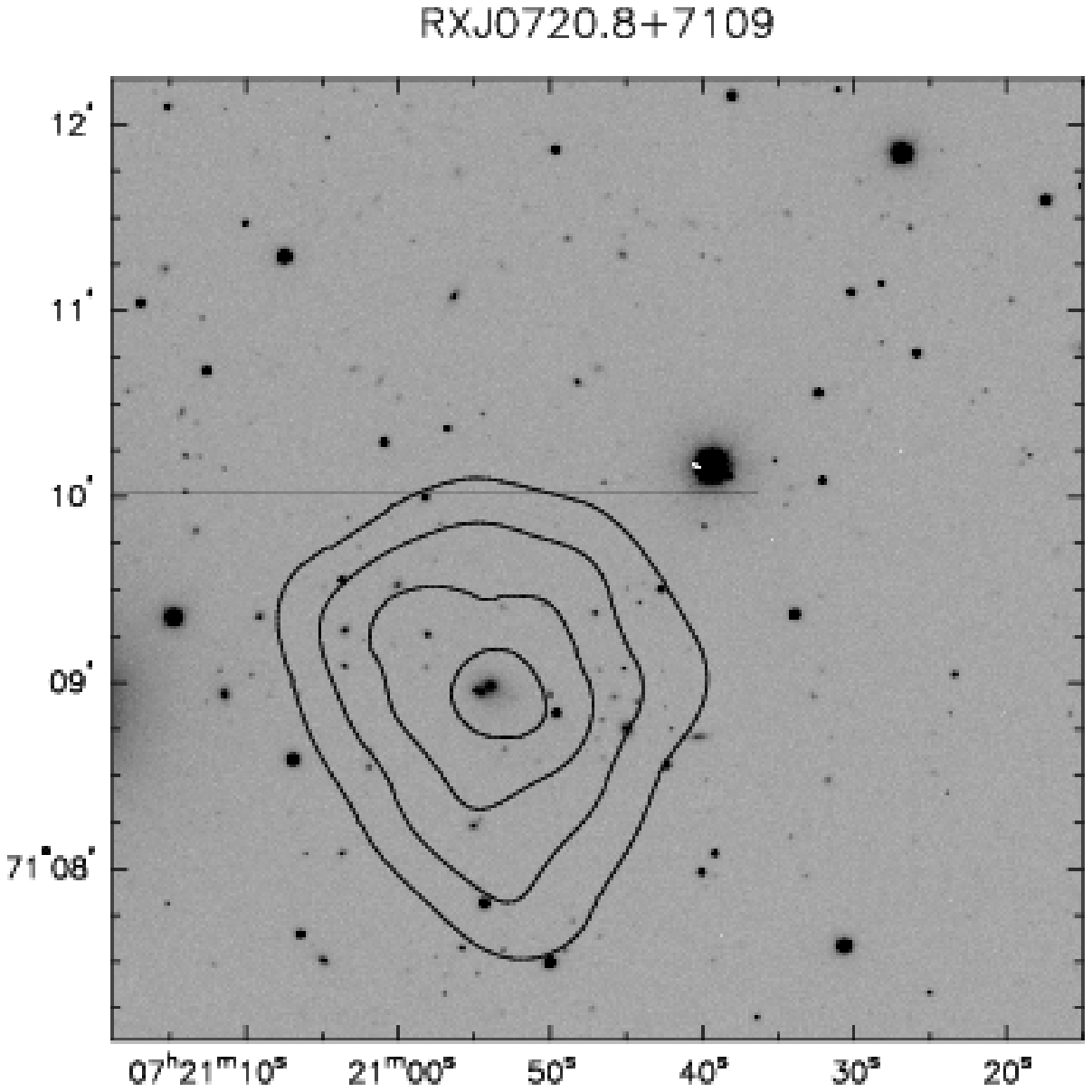}
\vskip 0.05cm
\includegraphics[width=2.5in]{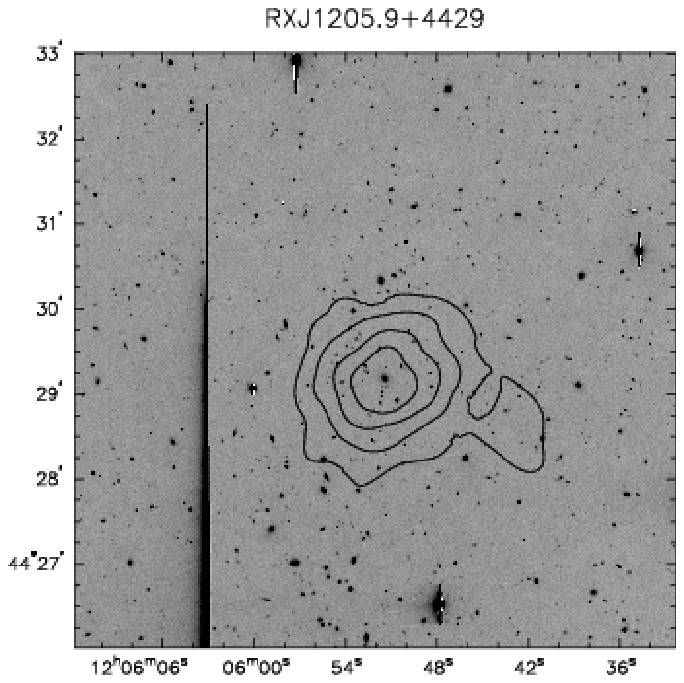}
\includegraphics[width=2.5in]{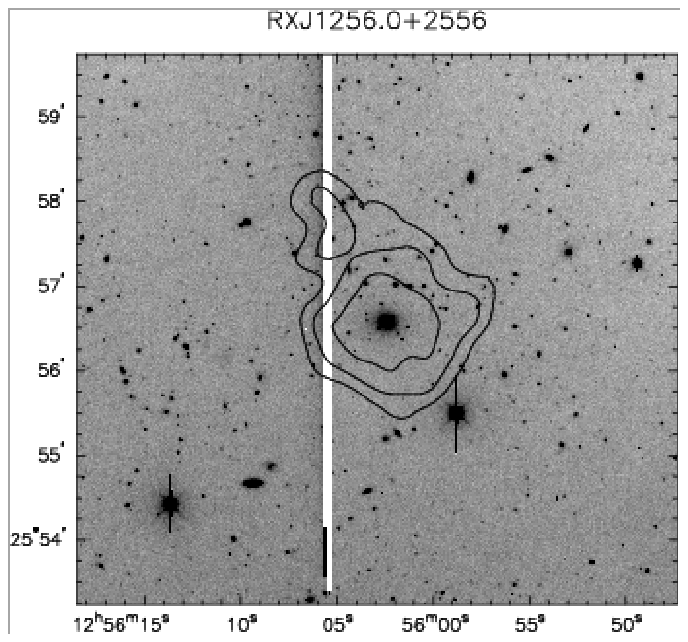}
\vskip 0.05cm
\includegraphics[width=2.5in]{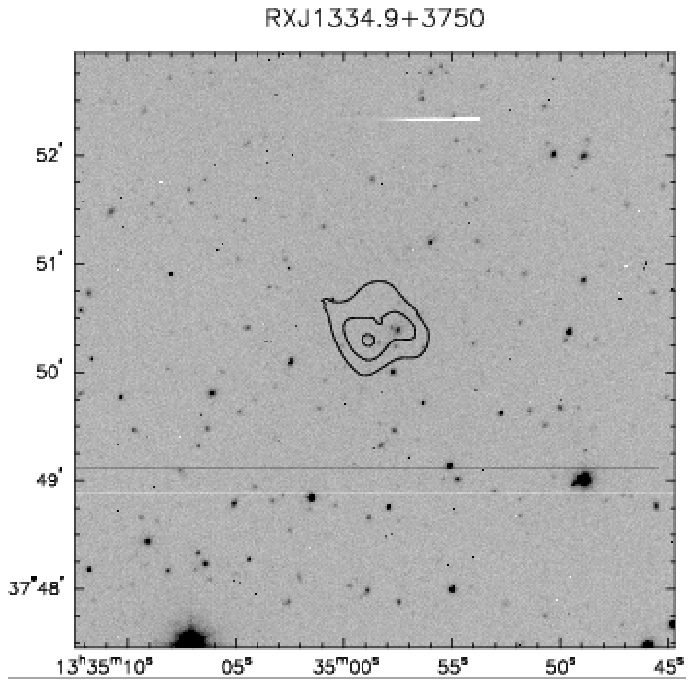}
\includegraphics[width=2.5in]{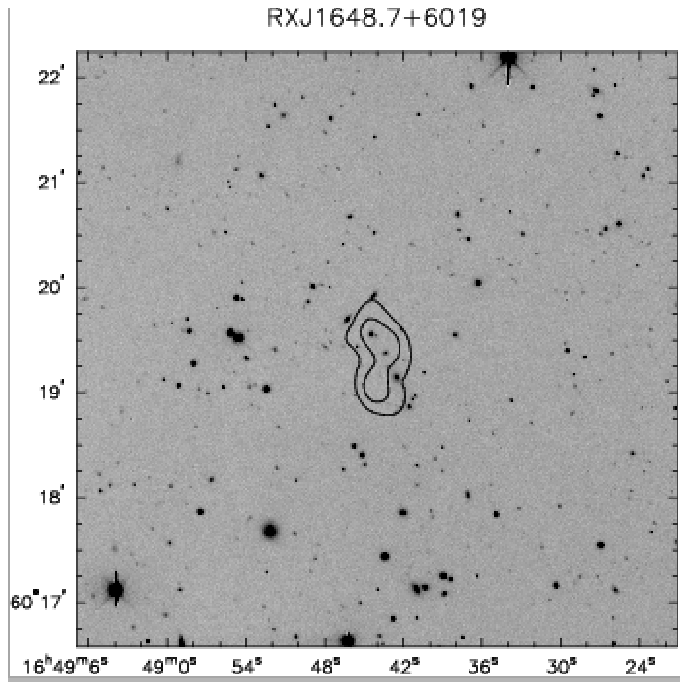}
\vskip 0.05cm
\noindent Figure 2 - \textit{XMM-Newton} contours of the diffuse X-ray
emission overlaid on optical images of the fields for the six systems
with \textit{XMM-Newton} data (Paper II).  The contours correspond to
5$\sigma$, 10$\sigma$, 20$\sigma$ and 40$\sigma$ above the background
level.

\vfill\eject

{\center
\includegraphics[angle=0,width=2.5in]{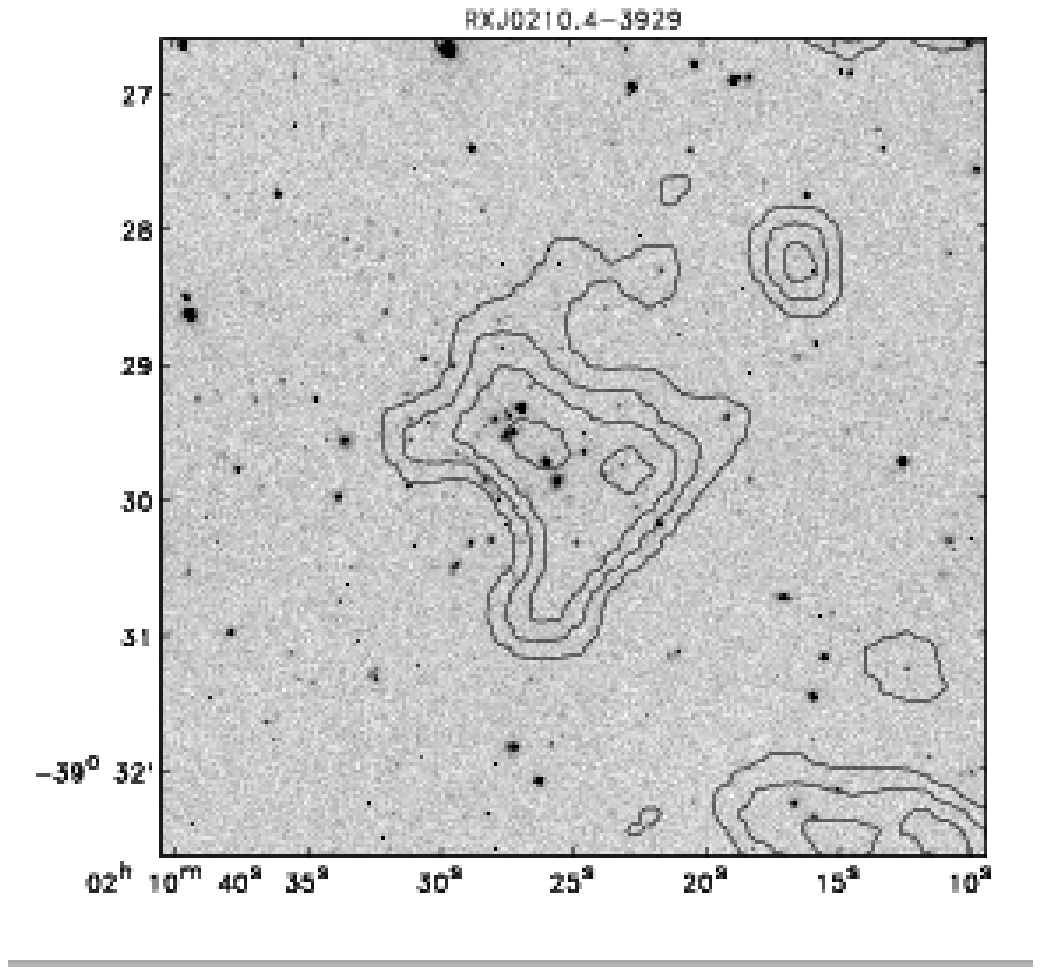}
\vskip 0.05cm
\includegraphics[angle=0,width=2.5in]{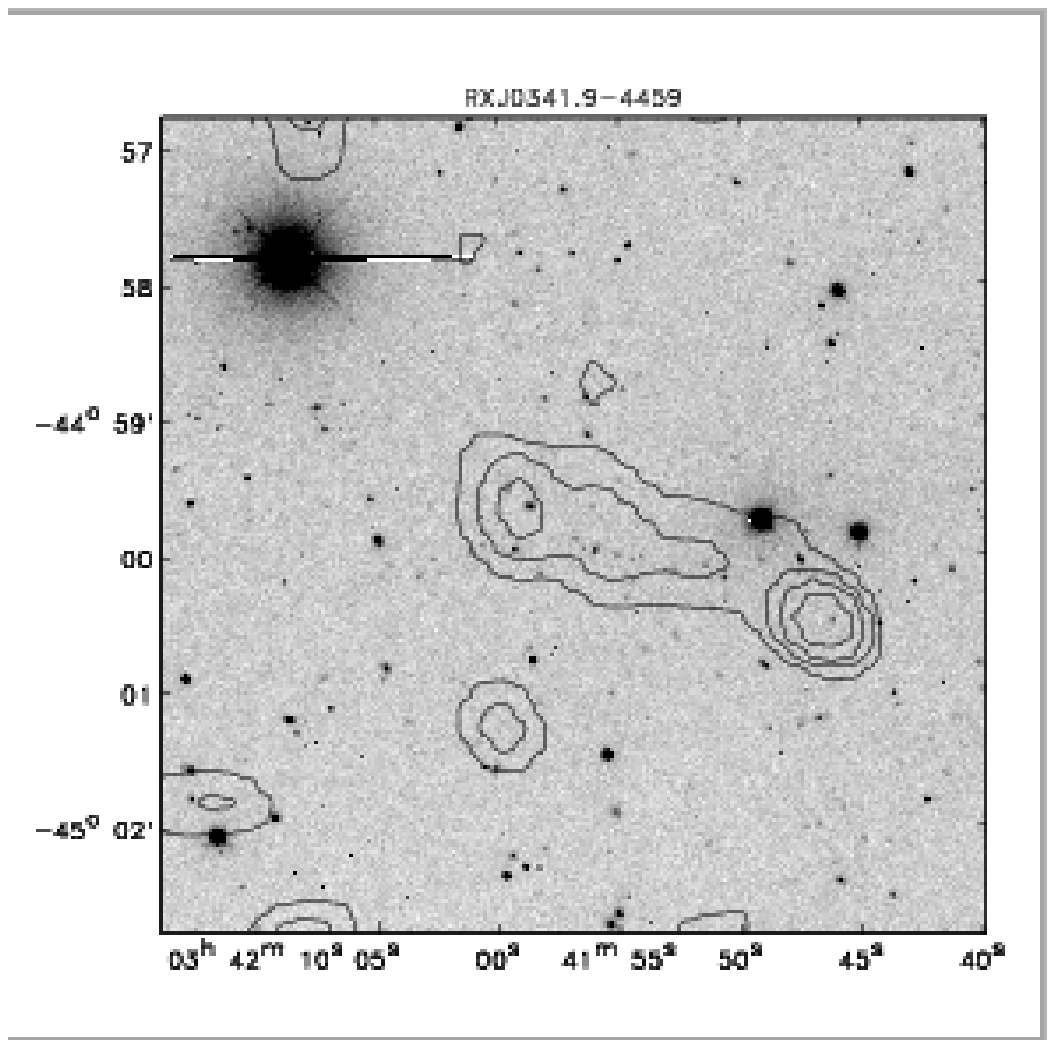}
\vskip 0.05cm
\includegraphics[angle=0,width=2.5in]{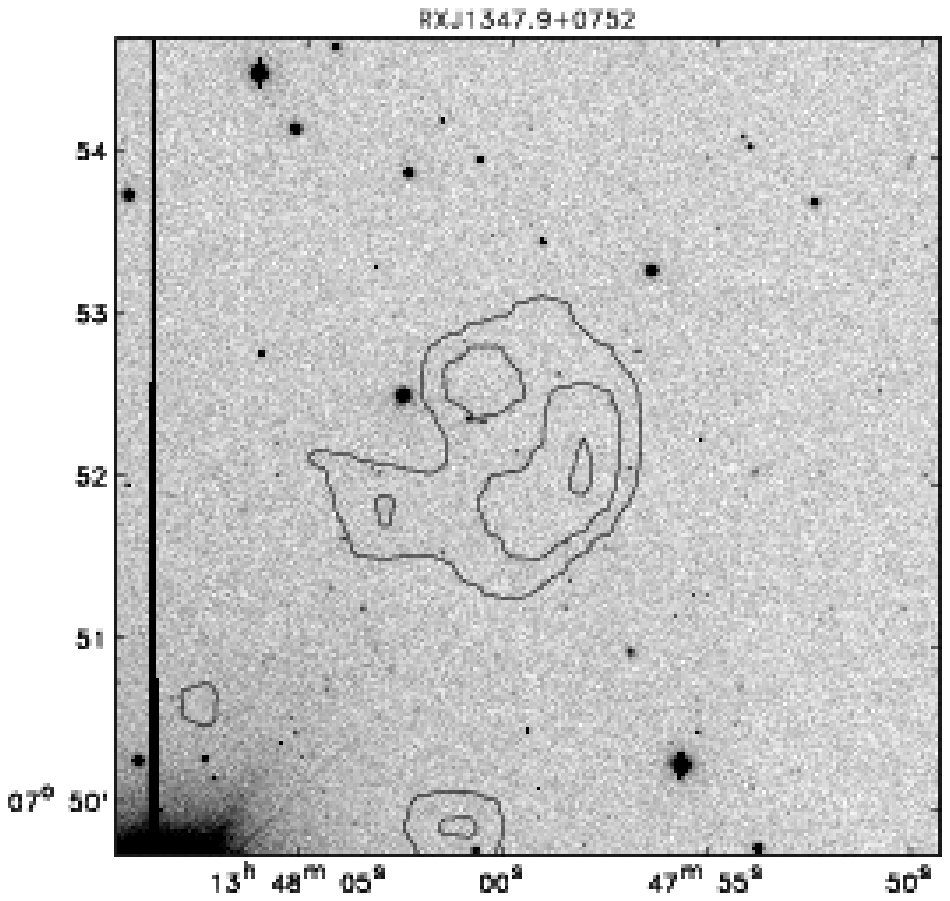}
\vskip 0.05cm
}
\noindent Figure 3 - \textit{ROSAT} contours taken from the RDCS
overlaid on optical images of the fields for the three systems in our
sample without \textit{XMM-Newton} data. The contours correspond to 
3$\sigma$,5$\sigma$,7$\sigma$,10$\sigma$ and 20$\sigma$ above the 
background level. 
                                                                                           
\vfill\eject

\includegraphics[angle=0,width=2.5in]{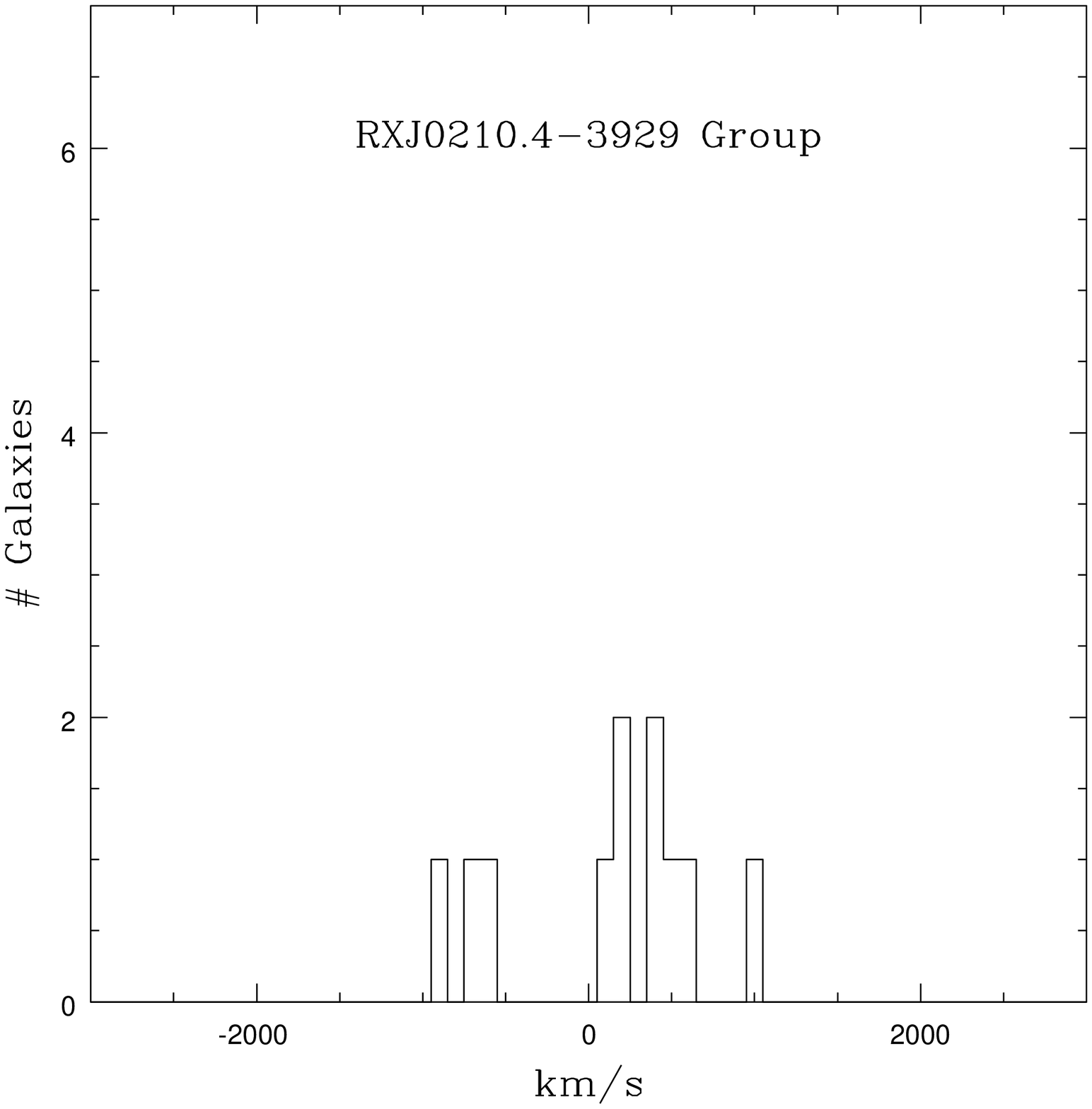}
\includegraphics[angle=0,width=2.5in]{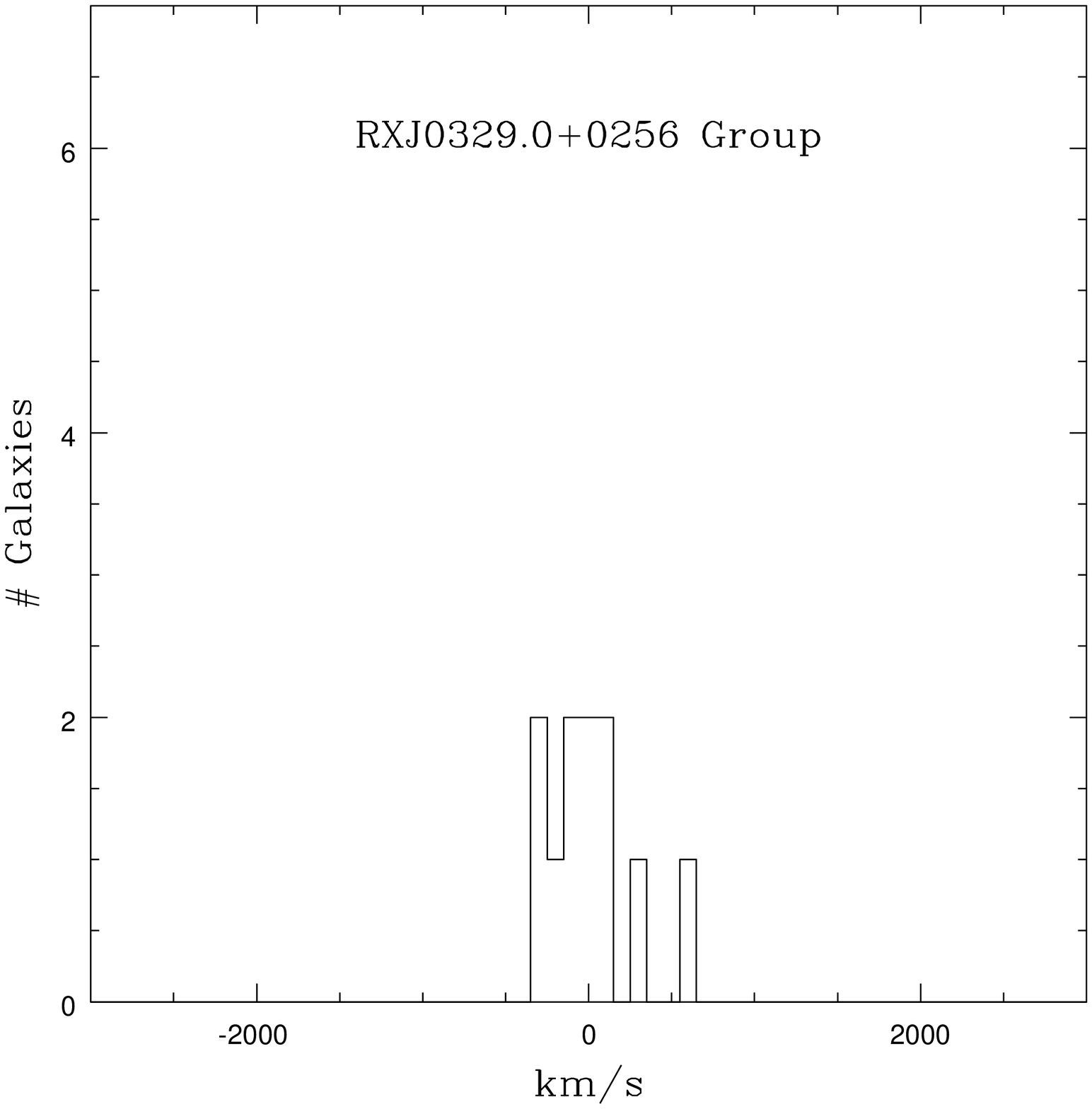}
\vskip 0.1cm
\includegraphics[angle=0,width=2.5in]{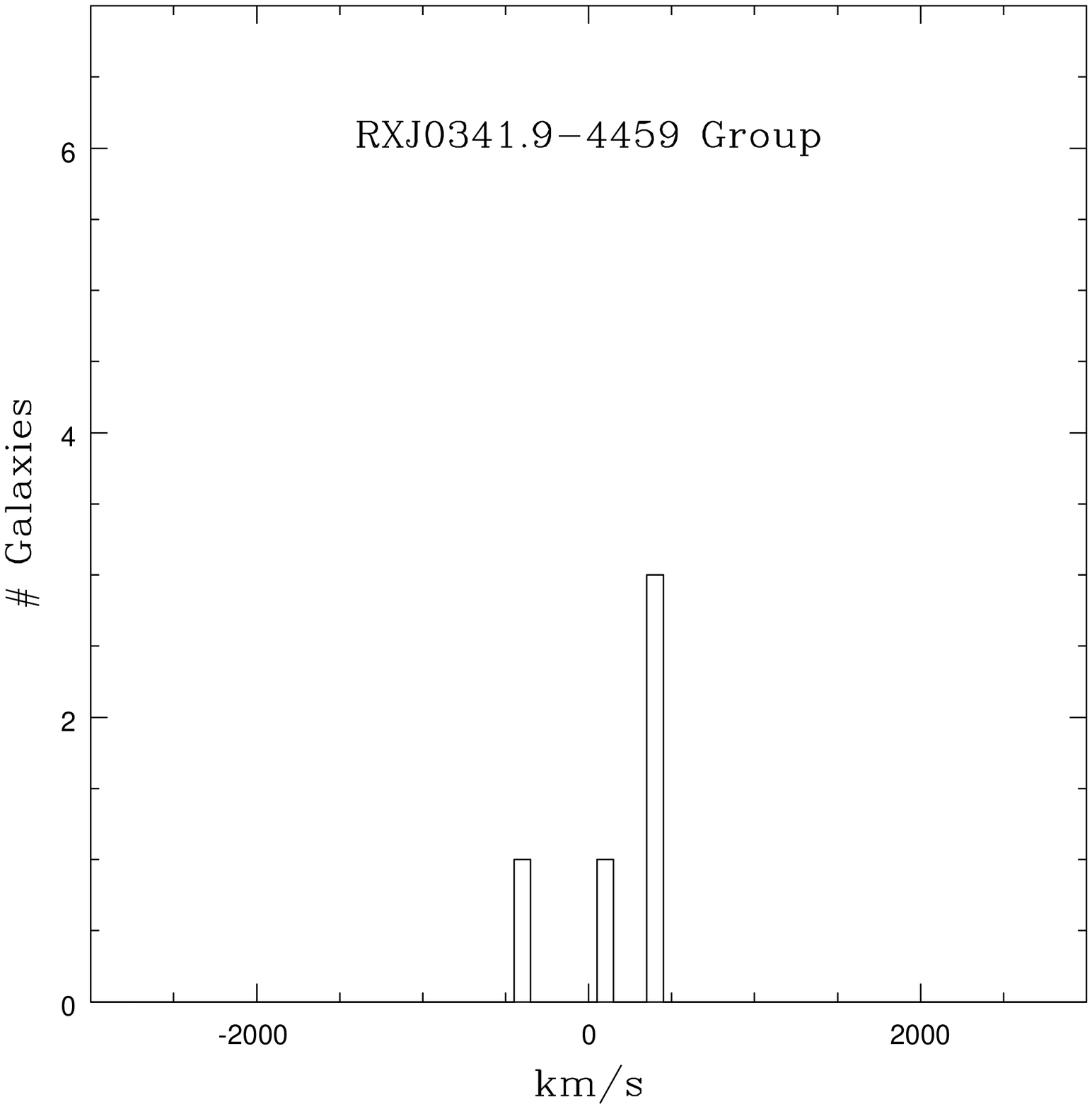}
\includegraphics[angle=0,width=2.5in]{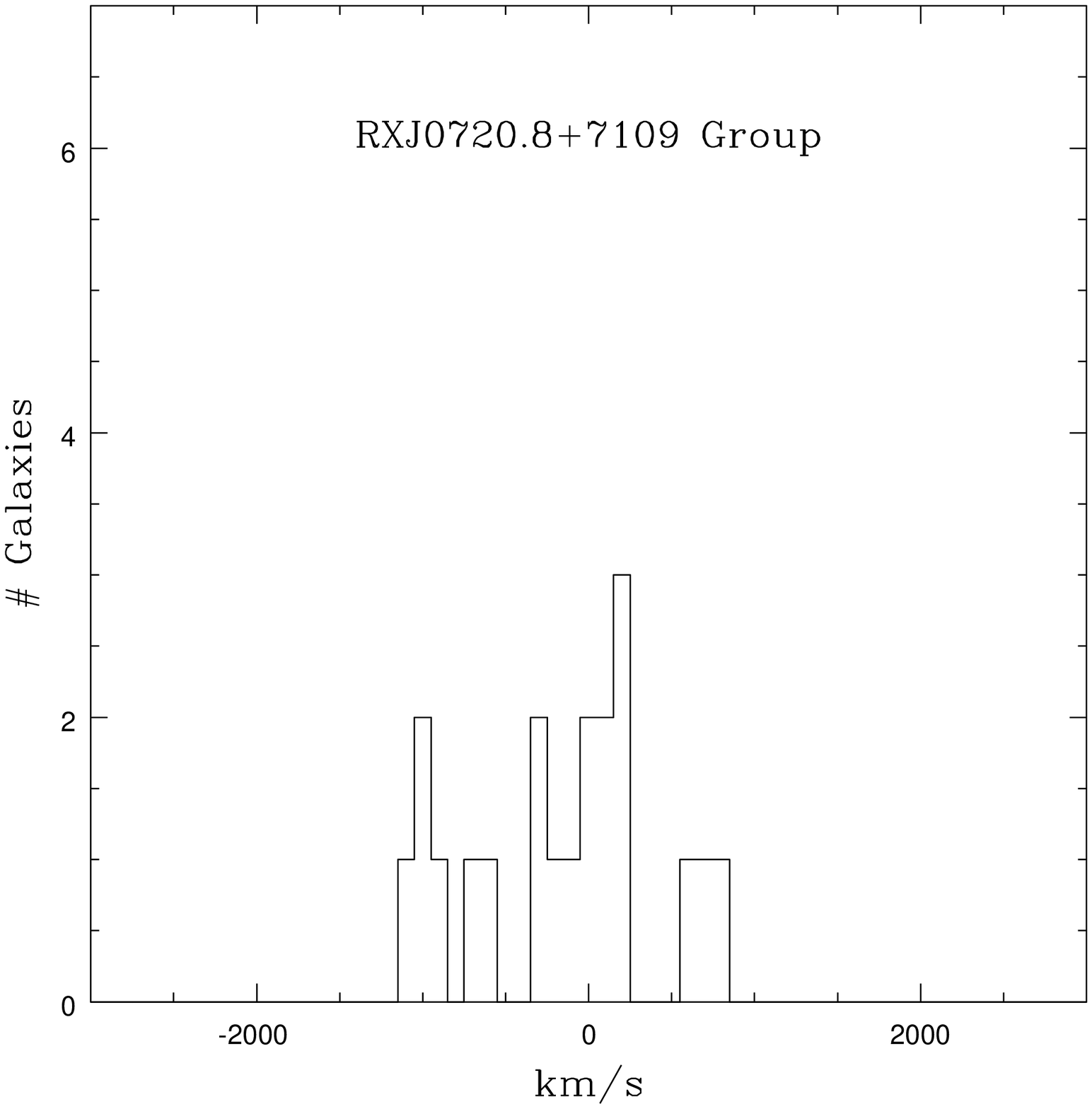}
\vskip 0.1cm
\includegraphics[angle=0,width=2.5in]{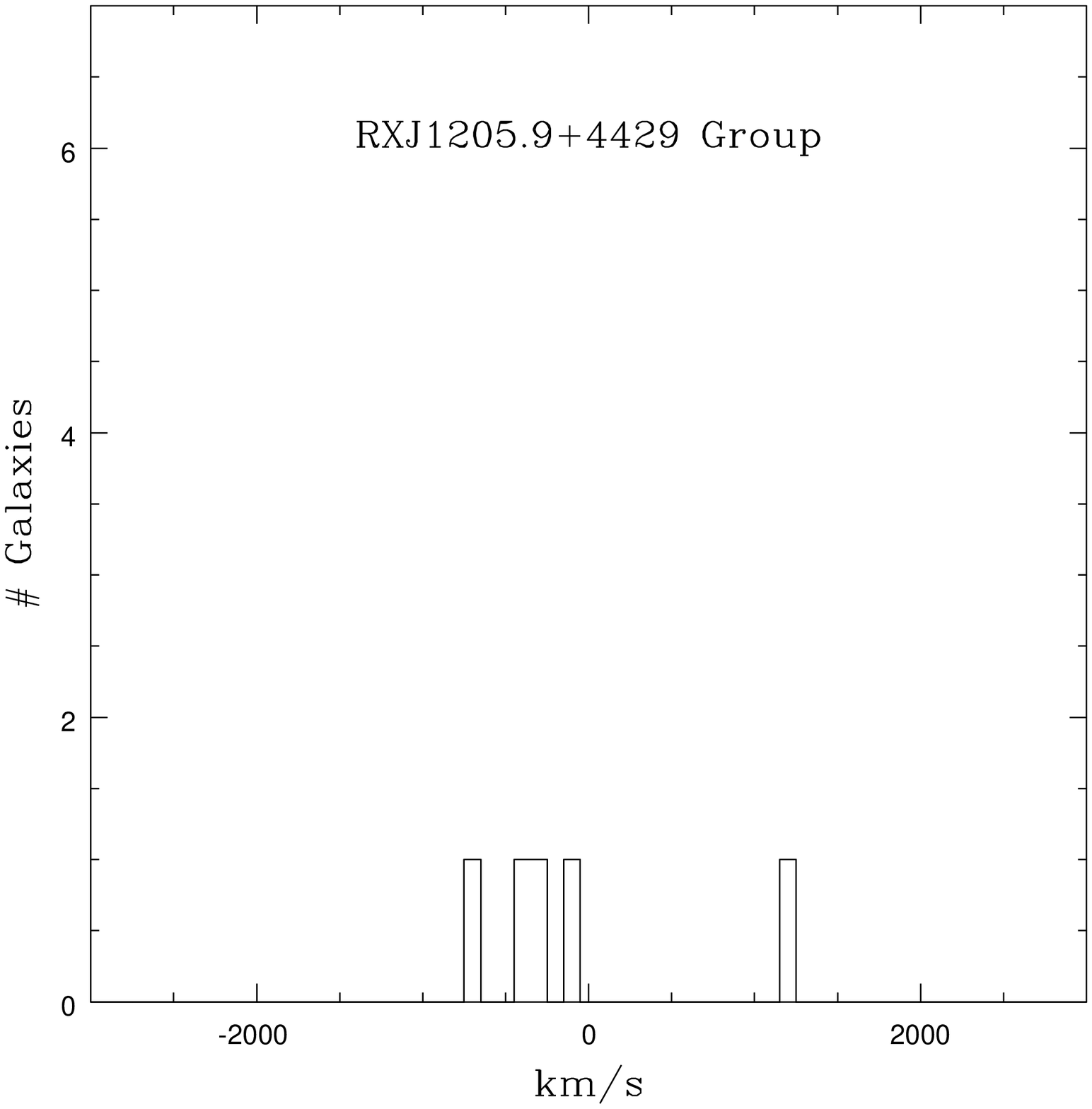}
\includegraphics[angle=0,width=2.5in]{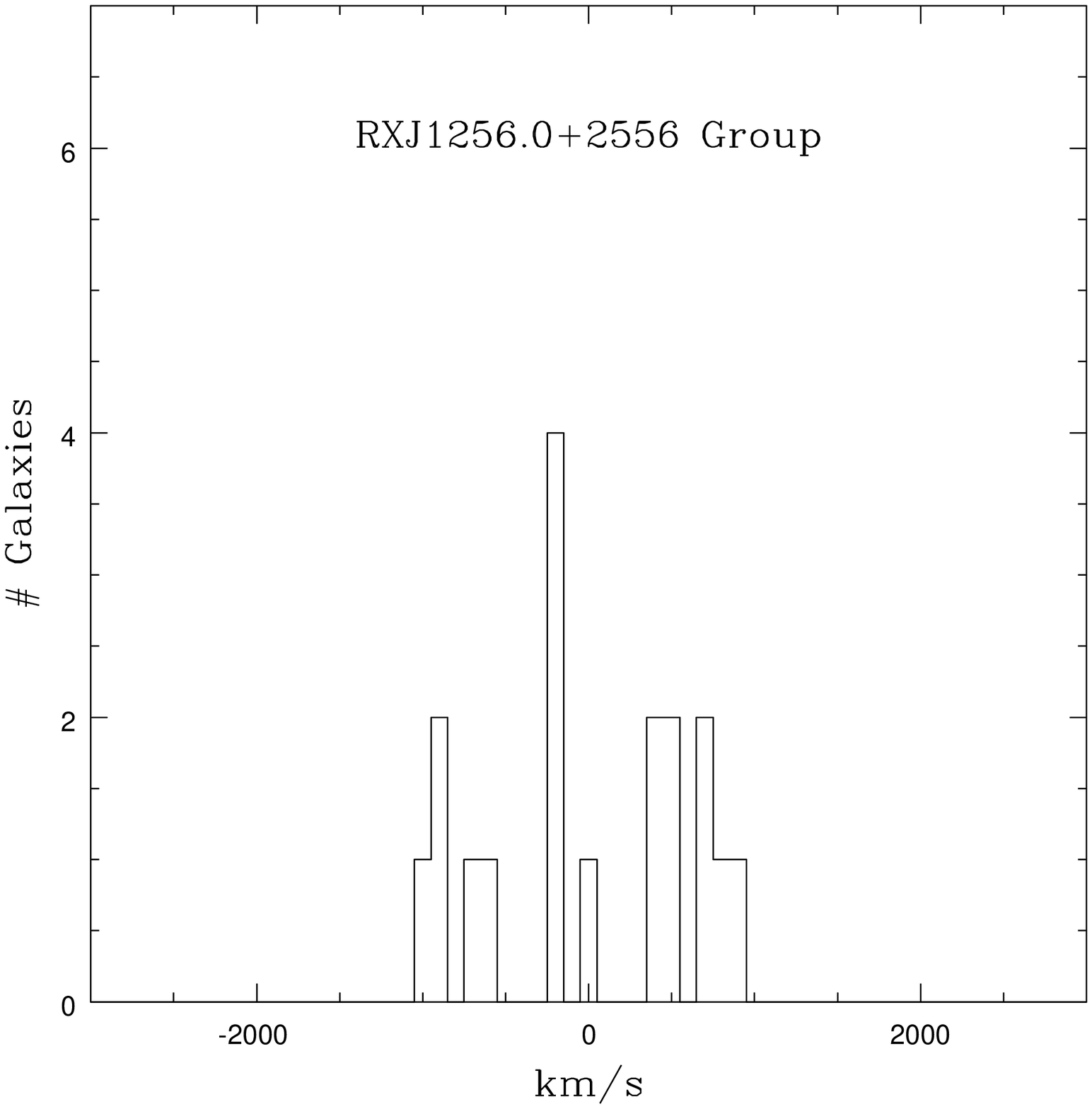}
\vskip 0.1cm
\includegraphics[angle=0,width=2.5in]{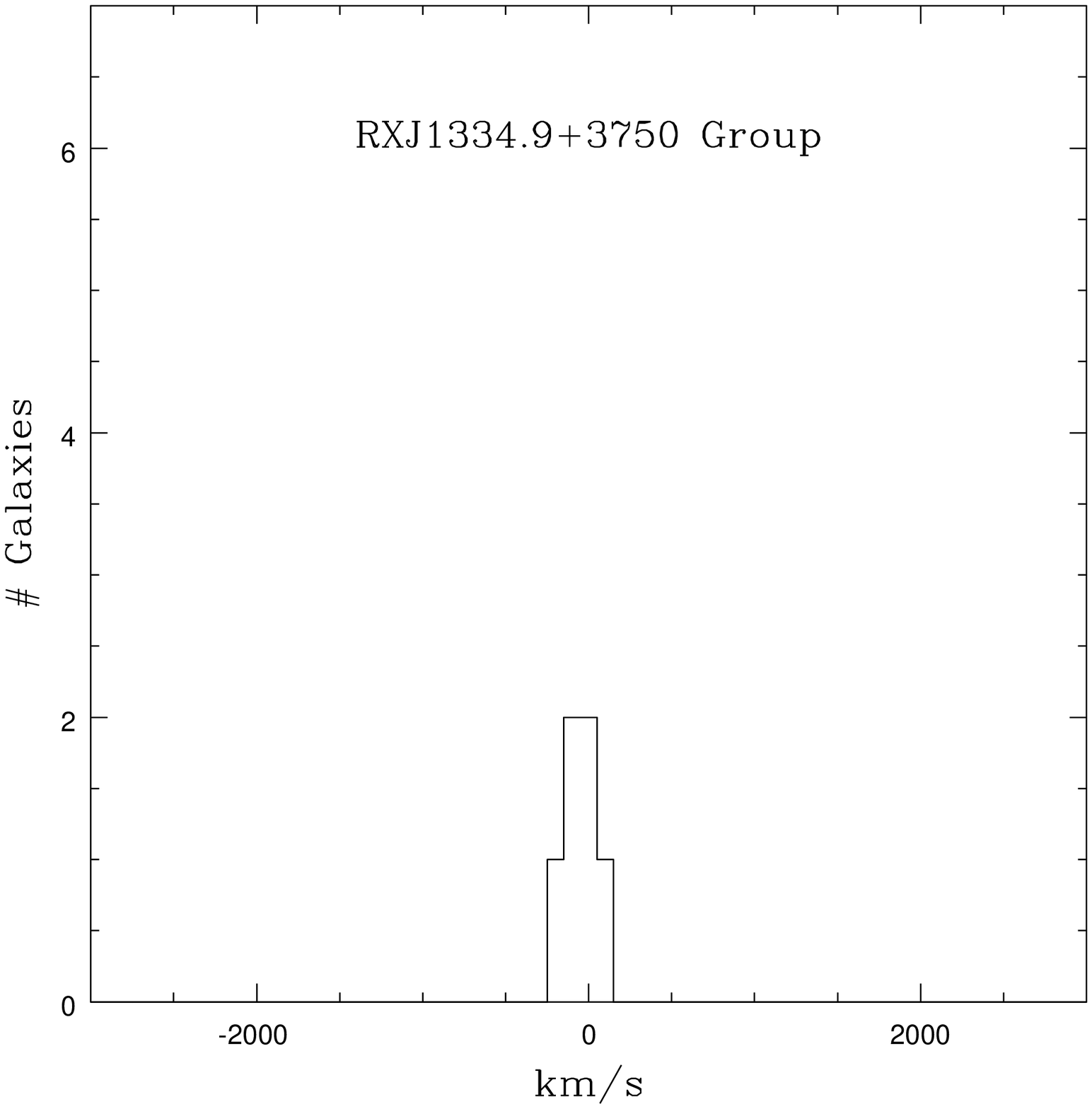}
\includegraphics[angle=0,width=2.5in]{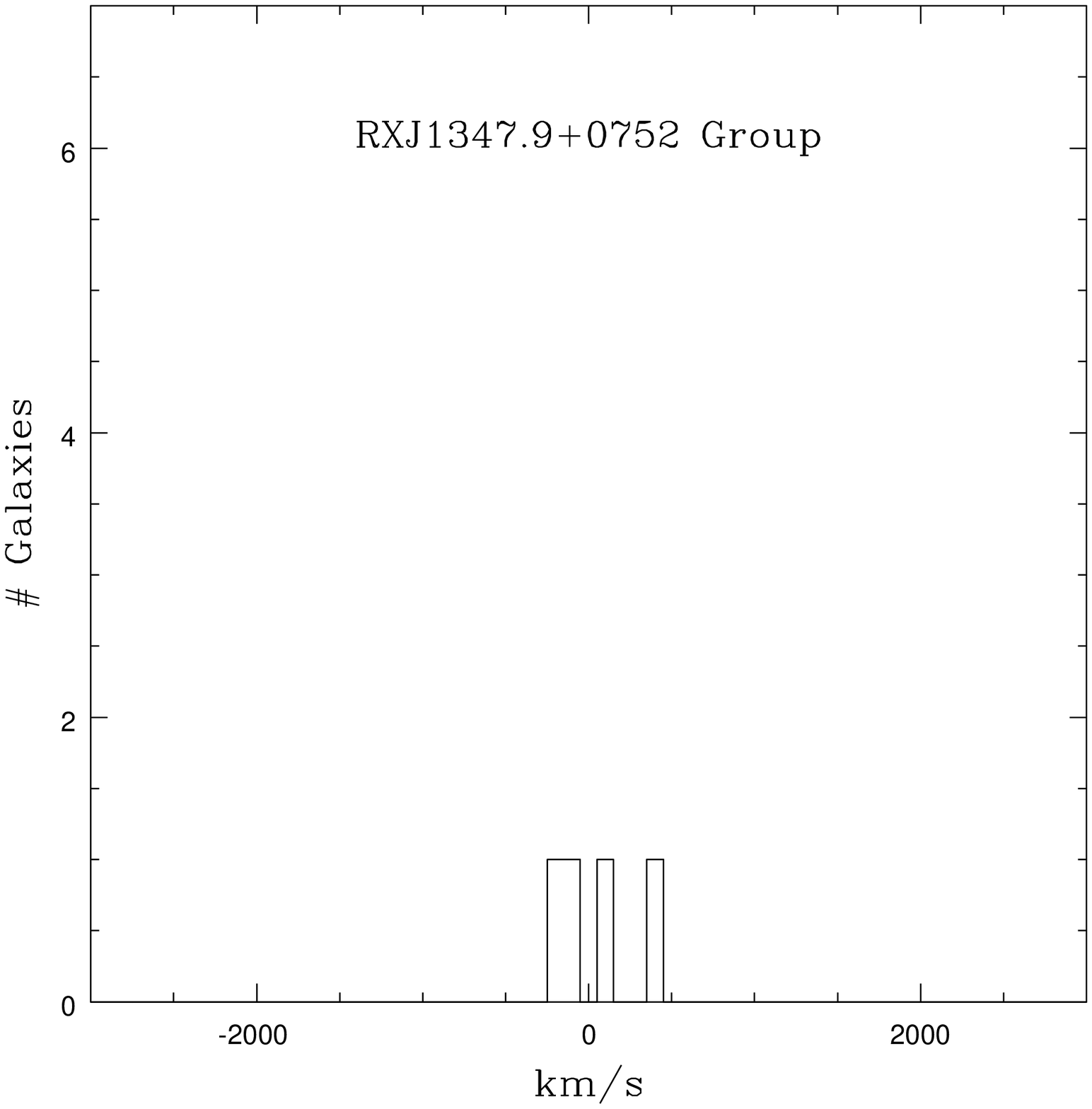}
\vskip 0.1cm
{\center
\includegraphics[angle=0,width=2.5in]{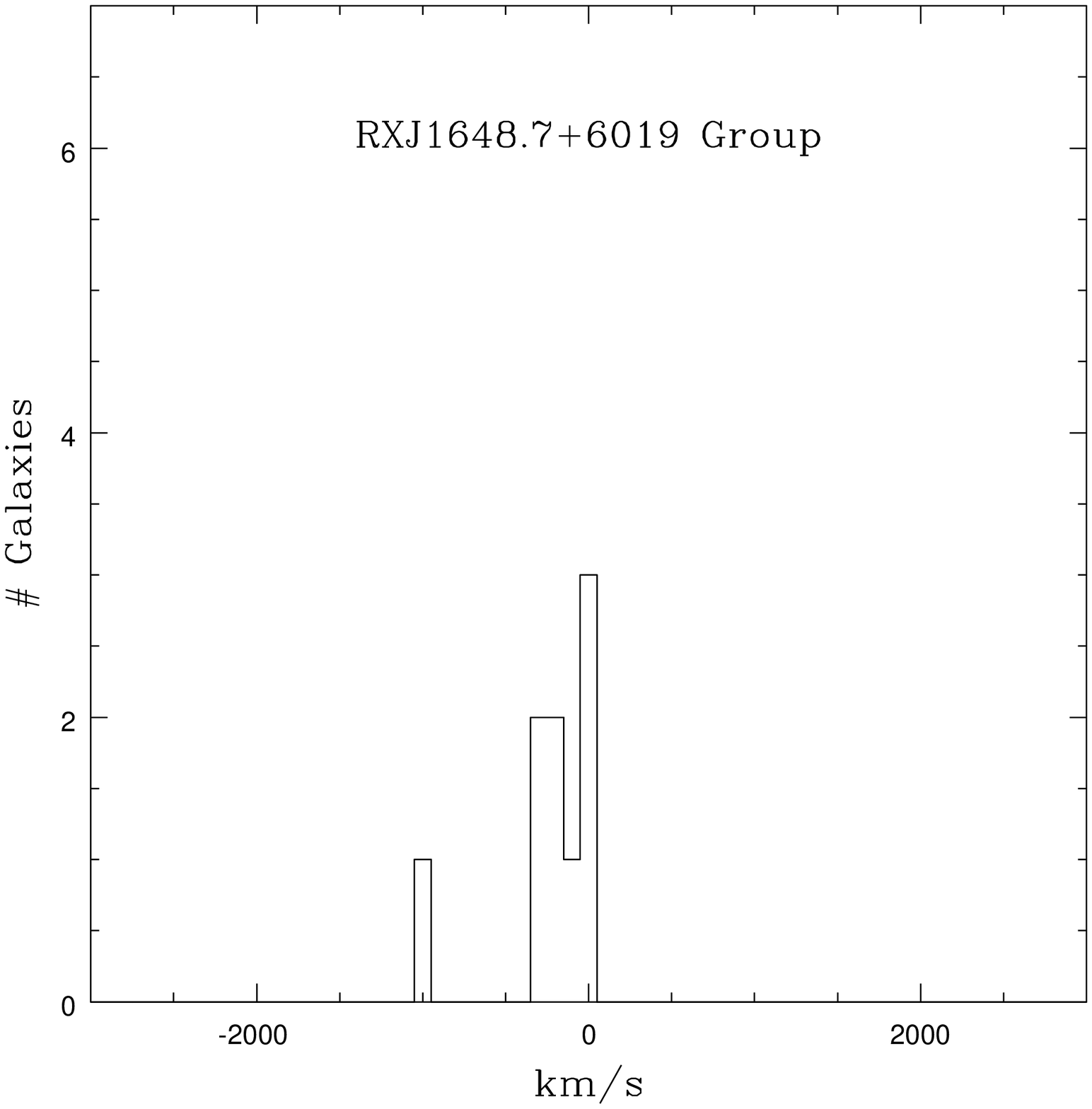}
\vskip 0.2cm
}

\noindent Figure 4 - Distribution of member velocities 
relative to the mean group velocity.

\vskip 6in

\includegraphics[width=5in]{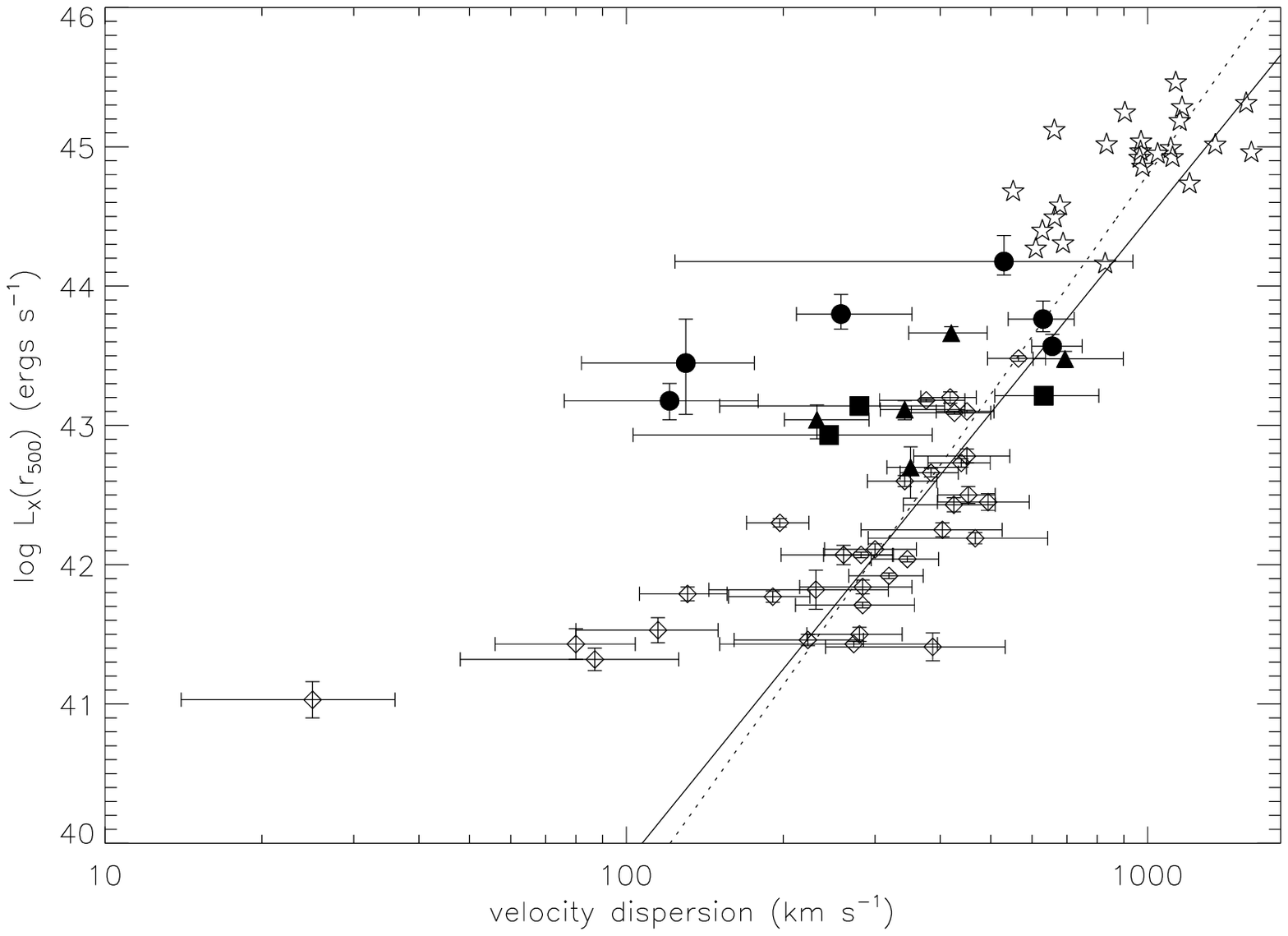}
\vskip 0.2cm
\noindent Figure 5 - Relationship between optical velocity dispersion
 and X-ray luminosity (L$_{\rm X}$) for the groups in the present
 sample with \textit{XMM-Newton} observations (filled circles), the
 groups in our sample with only \textit{ROSAT} observations (filled
 squares), the moderate redshift groups from \citet{wil05} (filled
 triangles) and the low redshift groups from \citet{osm04} (open
 diamonds). For the \textit{ROSAT}-only groups, a temperature of 2 keV
 has been assumed to estimate the X-ray luminosity. Also shown are 
 a fit to the low redshift GEMS groups (solid line) and  a fit to 
the Markevitch (1998) cluster sample (dotted line)(Helsdon \& Ponman, in preparation).

\includegraphics[width=5in]{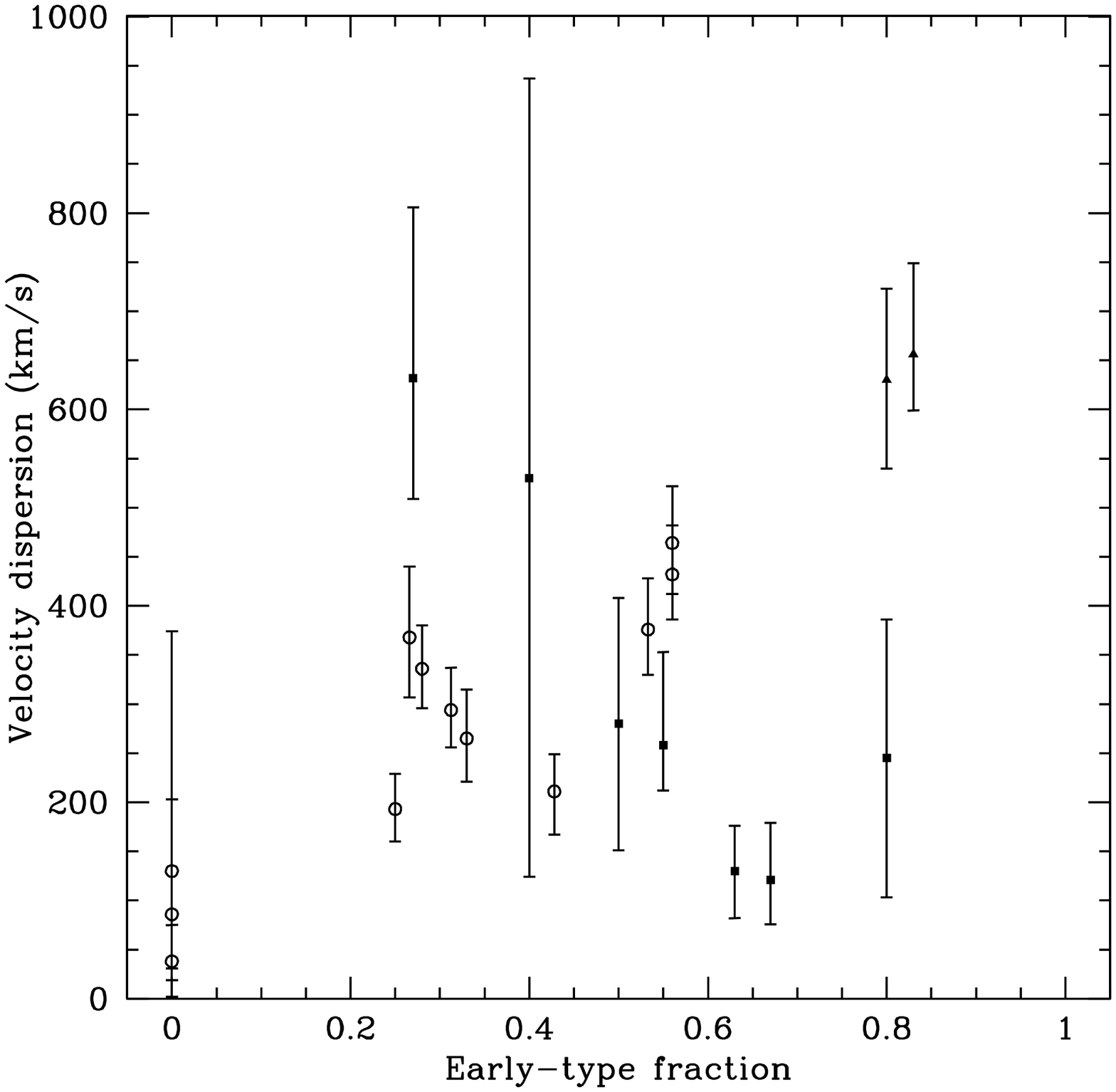}
\vskip 0.2cm
\noindent Figure 6 - Relationship between early-type fraction and
optical velocity dispersion for the present sample (filled points) and
the low redshift groups from \citet{zab98} (open circles).  The two
groups from the current sample with the best membership data are
plotted as triangles.  Note that these two systems appear to extend
the correlation found by \citet{zab98} out to systems with velocity
dispersions $\sim$ 600 km s$^{-1}$.

\includegraphics[width=2.5in]{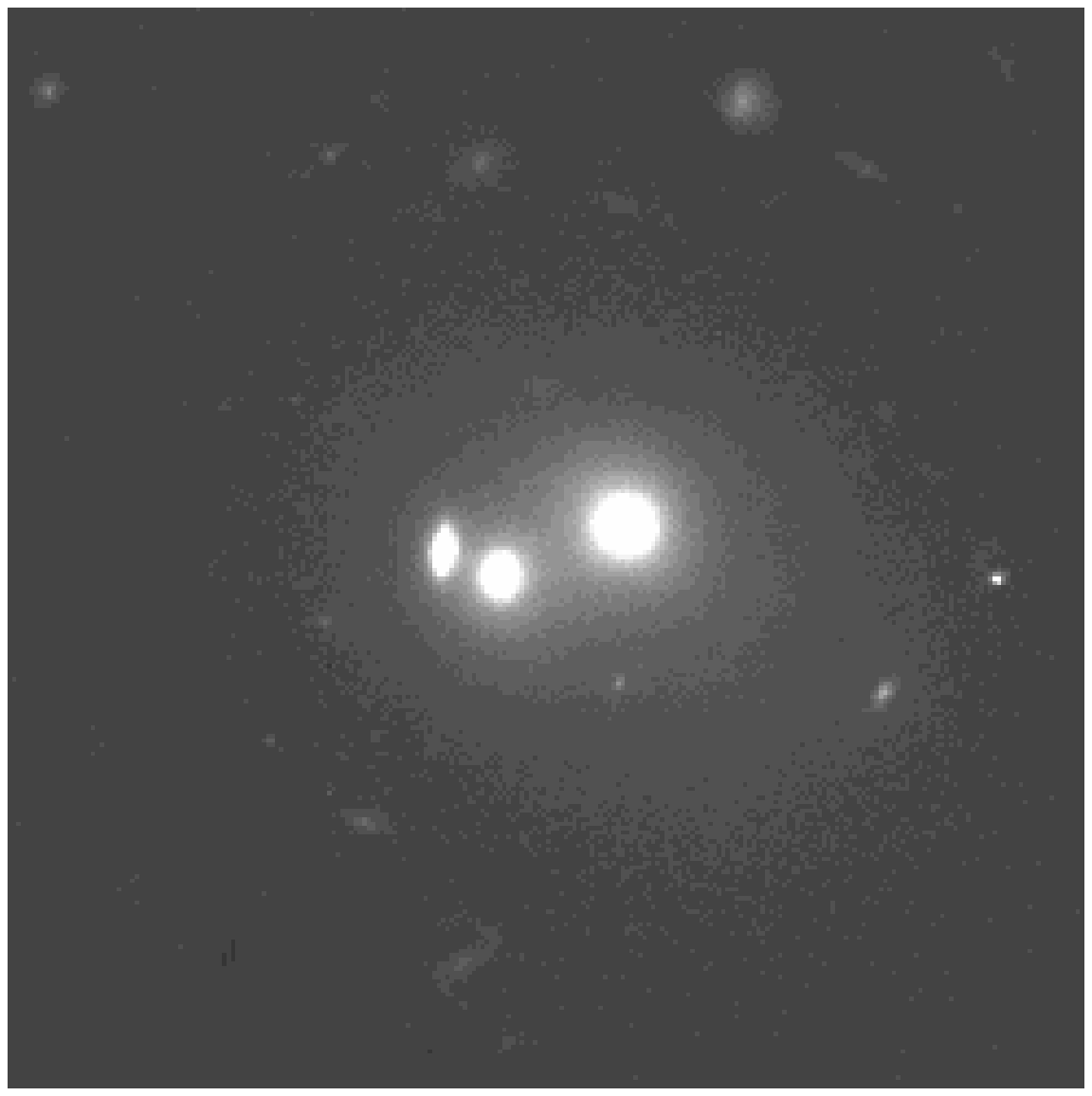}
\includegraphics[width=2.5in]{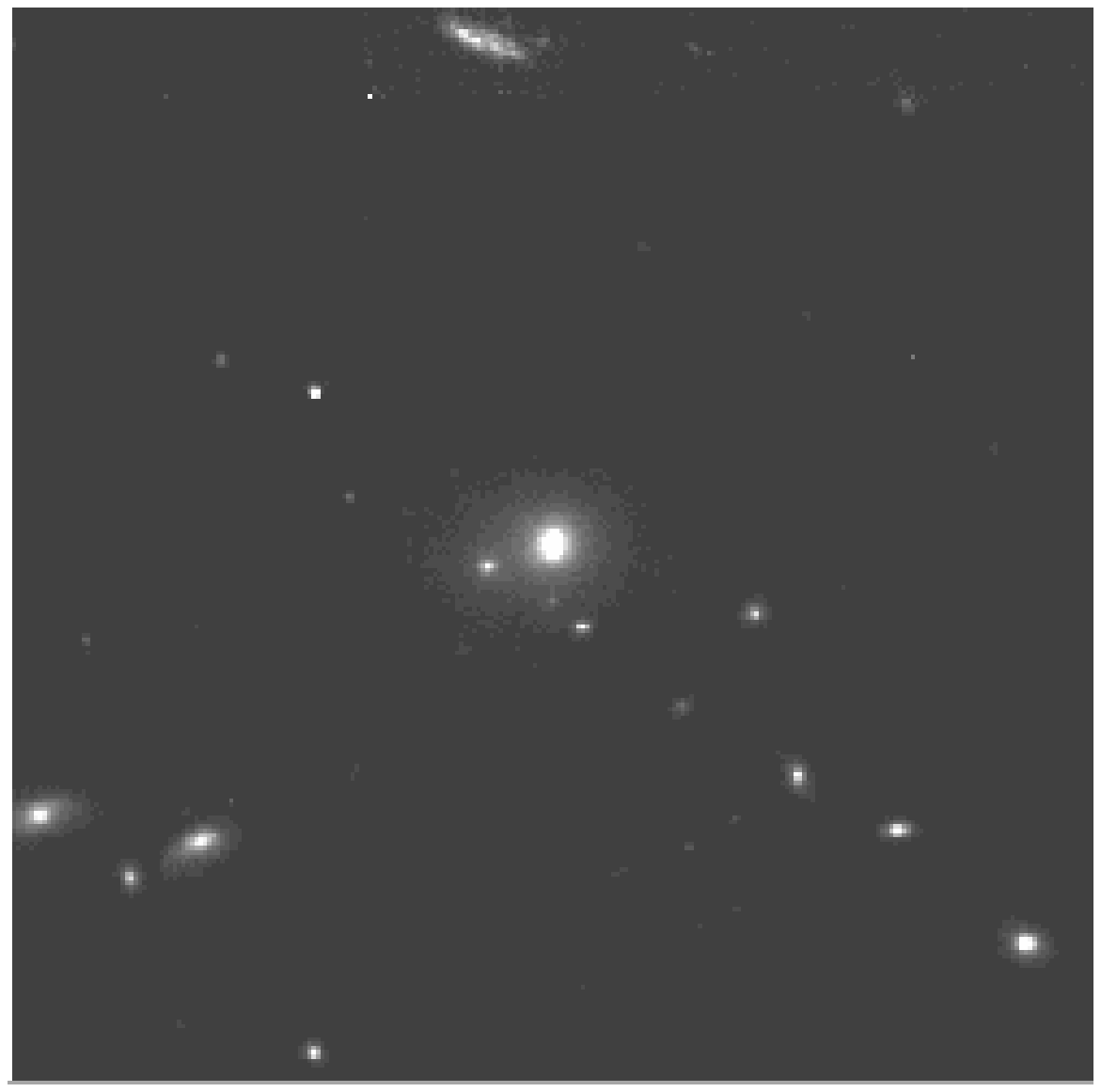}
\includegraphics[width=2.5in]{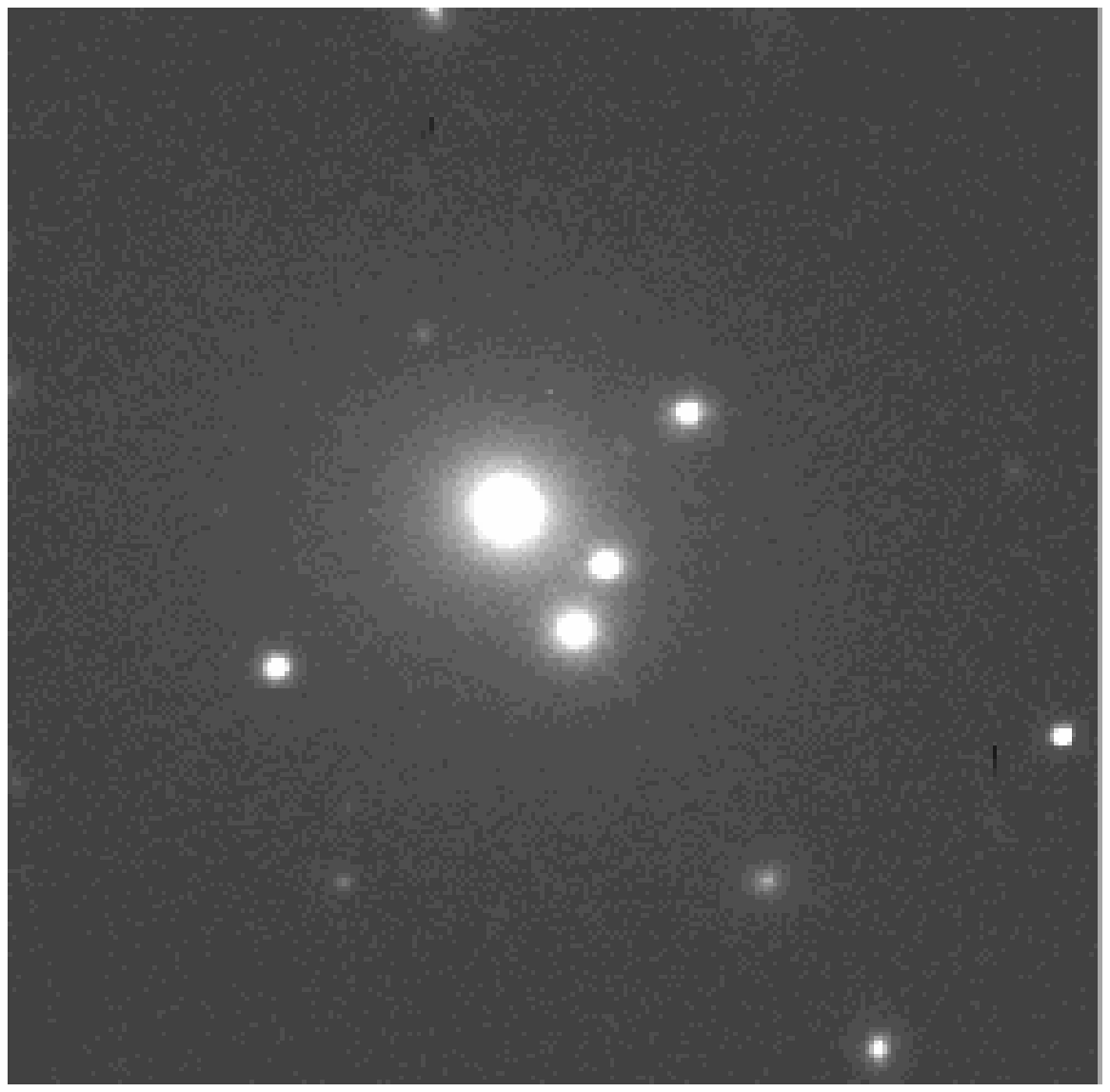}
\vskip 0.2cm
\noindent Figure 7 - \textit{HST WFPC2} images of the centers of the
RXJ0720.8+7109 (left), RXJ1205.9+4429 (middle) and RXJ1256.0+2556
(right) groups. The region plotted corresponds to a 200 kpc x 200 kpc
region in each case.



\begin{deluxetable}{lrrcrl}
\tablecolumns{6}
\tablewidth{0pc}
\tablecaption{Photometric and Spectroscopic Results}
\tablehead{
\colhead{RA (J2000)}   & \colhead{Dec (J2000)}    & \colhead{R mag.} &
\colhead{z}    & \colhead{z error} & \colhead{z type}}
\startdata
02 10 09.90 & -39 26 38.7 & 17.1& 0.1659 & 0.0006 & abs + em \\
02 10 10.78 & -39 20 19.3 & 19.1 & 0.1758 & 0.0004 & abs + em \\
02 10 13.39 & -39 32 58.9 & 17.1& 0.1665 & 0.0008 & abs + em \\
02 10 14.40 & -39 26 51.9 & 19.9& 0.3025 & 0.0003 & abs \\
02 10 14.98 & -39 32 42.6 & 17.8 & 0.1678& 0.0011 & abs \\
02 10 21.10 & -39 32 42.5 & 18.8 & 0.1643 & 0.0010 & abs + em \\
02 10 21.71 & -39 20 11.3 & 19.2 &  0.3702 & 0.0006 & abs \\
02 10 24.52 & -39 29 39.4 & 19.2 & 0.3075 & 0.0008 & abs \\
\enddata
\tablecomments{The full table is available electronically.}
\end{deluxetable}

\begin{deluxetable}{lrrrclcl}
\rotate{}
\tablecolumns{8}
\tablewidth{0pc}
\tablecaption{Group Properties}
\tablehead{
\colhead{Group} & \colhead{RA (J2000)}   & \colhead{Dec (J2000)}    & \colhead{redshift} &
\colhead{Members}    & \colhead{$\sigma$$_{\rm biwt}$} & \colhead{Early-type Fraction} & \colhead{Comments}}\startdata
RXJ0210.4-3929 & 02 10 24.89 & -39 29 37.1 & 0.3058 & 11 & 632$^{+174}_{-123}$ & 0.27 & No early-type
 BGG; spiral
dominated \\
RXJ0329.0+0256 & 03 29 02.82  & +02 56 25.2 & 0.4122 & 11 & 258$^{+95}_{-46}$& 0.55 & BGG at
group center \\
RXJ0341.9-4459 & 03 41 56.83 & -44 59 46.7 & 0.4063 & 5  &  245$^{+141}_{-142}$ & 0.80 &
BGG possibly offset in velocity \\
RXJ0720.8+7109 & 07 20 54.04 & +71 08 57.9& 0.2309 & 20 & 630$^{+93}_{-90}$& 0.80 &
BGG with three components \\
RXJ1205.9+4429 & 12 05 51.44 & +44 29 11.0 & 0.5926 & 5  & 530$^{+407}_{-406}$ & 0.40 &
BGG with two components \\
RXJ1256.0+2556 & 12 56 02.34 & +25 56 37.1 & 0.2316 & 18 & 656$^{+93}_{-57}$
&0.83 & BGG with three components\\
RXJ1334.9+3750 & 13 34 58.95 & +37 50 15.7 & 0.3839 & 6  & 121$^{+58}_{-45}$&0.67 &
BGG offset in X-rays and velocity \\
RXJ1347.9+0752 & 13 47 59.51 & +07 52 12.1 & 0.4649 & 4  & 280$^{+128}_{-129}$ &0.50 &
BGG possibly offset in velocity \\
RXJ1648.7+6019 & 16 48 43.63& +60 19 21.5 & 0.3763 & 8  & 130$^{+46}_{-48}$ & 0.63 &
No dominant BGG; brightest early-type offset in X-rays and velocity
 \\
\enddata
\end{deluxetable}


\end{document}